\documentclass[12pt]{article}
\usepackage{axodraw}
\usepackage{epsfig}   
\usepackage{amssymb}
\usepackage{latexsym}
\usepackage{amsmath}

\newcommand{\ul} {\tilde u_L}
\newcommand{\ur} {\tilde u_R}
\newcommand{\dl} {\tilde d_L}
\newcommand{\dr} {\tilde d_R}
\newcommand{\ulb} {\bar{\tilde u}_L}
\newcommand{\urb} {\bar{\tilde u}_R}
\newcommand{\dlb} {\bar{\tilde d}_L}
\newcommand{\drb} {\bar{\tilde d}_R}

\newcommand{\gsim}{\raisebox{-0.13cm}{~\shortstack{$>$ \\[-0.07cm] $\sim$}}~}
\newcommand{\eqa} {\begin{eqnarray} }
\newcommand{\eqe} {\end{eqnarray}}
\newcommand{\beq} {\begin{equation}}
\newcommand{\eeq} {\end{equation}}

\begin{document}
\pagestyle{empty}
\begin{flushright}
KIAS--P07049 \\
September 2007
\end{flushright}
\begin{center}
{\large\sc {\bf Electroweak Contributions to Squark Pair Production at the
    LHC}} 

\vspace{1cm}
{\sc Sascha Bornhauser$^{1,2}$, Manuel Drees$^{1,2}$, Herbi K. Dreiner$^1$}
and {\sc Jong Soo Kim$^1$} 

\vspace*{5mm}
{}$^1${\it Physikalisches Institut, Universit\"at Bonn, Nussallee 12, D53115
  Bonn,  Germany} \\
{}$^2${\it KIAS, School of Physics, Seoul 130--012, Korea}
\end{center}
\vspace*{1cm}
\begin{abstract}
  In this paper we compute electroweak contributions to the production of
  squark pairs at hadron colliders. These include the exchange of electroweak
  gauge bosons in the $s-$channel as well as electroweak gaugino exchange in
  the $t-$ and/or $u-$channel. In many cases these can interfere with the
  dominant QCD contributions. As a result, we find sizable contributions to
  the production of two $SU(2)$ doublet squarks. At the LHC, they amount to 10
  to 20\% for typical mSUGRA (or CMSSM) scenarios, but in more general
  scenarios they can vary between $-40$ and $+55\%$, depending on size and
  sign of the $SU(2)$ gaugino mass. The electroweak contribution to the total
  squark pair production rate at the LHC is about 3.5 times smaller.

\end{abstract}
\newpage
\setcounter{page}{1}

\pagestyle{plain}
\section{Introduction}

The Standard Model (SM) of particle physics, while very successful, suffers
from several theoretical as well as phenomenological problems. Many of these
can be solved by postulating the existence of superpartners \cite{susyrev} to
all known elementary particles, with masses at or below the TeV scale. This
solves the (technical part of the) hierarchy problem by stabilizing scalar
masses -- in particular, the mass of the SM Higgs boson -- against
quadratically divergent quantum corrections \cite{hierarchy}. Superstring
theory, which is our currently most promising avenue towards a perturbative
theory of quantum gravity, also requires supersymmetry (albeit not necessarily
at the TeV scale).

The simplest possibly realistic supersymmetric model is known as the Minimal
Supersymmetric extension of the Standard Model (MSSM). In addition to the
virtues described above, it also allows for one--step unification of the three
gauge couplings of the SM \cite{susyuni}. Moreover, it has several
phenomenological advantages. It can naturally accommodate \cite{susydm} the
existence of Dark Matter, which probably accounts for $\sim 80\%$ of all
matter in the Universe \cite{wmap}. It easily provides \cite{susygmu} the
additional quantum correction to the anomalous magnetic moment of the muon
that seems to be required by current data \cite{gmuexp}, if the SM prediction
\cite{gmuth} based on $e^+ e^- \rightarrow$ hadrons data can be trusted.
Finally, recent fits to precision electroweak data indicate a slight
preference of the MSSM over the SM \cite{elw}.

The idea that superparticles exist near the TeV scale will be tested
decisively at the LHC \cite{tdrs}. Here the production of squark pairs is
expected to play an important role. This final state can be accessed via
leading order strong interactions, i.e. the cross section is ${\cal
  O}(\alpha_s^2)$. Moreover, there are contributions with two valence quarks
in the initial state. Since valence quarks have the hardest of all parton
distribution functions (pdf's), i.e. show the slowest fall--off as Bjorken$-x
\rightarrow 1$, very massive quark pairs can be produced with appreciable
rate. For example, the leading order (LO) QCD cross section for the production
of degenerate first and second generation squarks with mass 1 TeV still
exceeds 0.5 pb, leading to 5,000 events per year even at low luminosity. One
can thus expect that the total cross section for squark pair production will
eventually be measured with a statistical uncertainty of a few
percent. Accurate predictions for this quantity are therefore obviously of
great interest.

The leading order QCD cross section for squark pair production at hadron
colliders has first been calculated in the 1980s \cite{qcdlo}. The NLO QCD
corrections were calculated about ten years ago \cite{qcdnlo}. The remaining
uncertainty from yet higher order QCD corrections should be at the ten percent
level. 

In this paper we compute the complete leading order electroweak contributions
to squark pair production at hadron colliders. Since in many cases
interference with QCD amplitudes is possible, this yields contributions of
${\cal O}(\alpha_S \alpha_W)$ as well as of ${\cal O}(\alpha_W^2$), where
$\alpha_W$ is a weak gauge coupling. We find that these change the total cross
section by only a few percent if at least one of the produced squarks is an
$SU(2)$ singlet. On the other hand, the cross section for the production of
two $SU(2)$ doublet squarks is changed by 10 to 20\% in typical mSUGRA
\cite{msugra} (or CMSSM) scenarios \cite{sps}; in scenarios without gaugino
mass unification \cite{code} the corrections can exceed
50\%.\footnote{Electroweak contributions to squark production have recently
  also been computed in ref.\cite{klasen}. However, the emphasis of that paper
  is on flavor observables; the impact on the total cross section, which is
  our main focus, is not discussed there. Ref.\cite{turks} also computes
  electroweak contributions to squark pair production. However, their emphasis
  is on tests of CP violation; besides, the analytical results in that paper
  seem to contain several errors (e.g., non--vanishing interference between
  gluon and photon exchange in the $s-$channel).} These new contributions peak
at small transverse momentum of the produced squark; they can therefore not be
subsumed in a constant ``$k-$factor''.

The remainder of this article is organized as follows. In the following
Section we give explicit expressions for the squared amplitudes for all
processes with two quarks in the initial state and two squarks in the final
state, where anti--particles are included. In Sec.~3 we present numerical
results for the total cross section as well as the $p_{T,\tilde q}$
distribution, before concluding in Sec.~4. A list of employed couplings is
given in the Appendix.

\section{Formalism}

In this section we present analytical results for the leading--order
parton--level squared matrix elements for the production of two (anti--)squarks
from two (anti--)quarks in the initial state. We will first give a `master
formula' in terms of general $s-, \, t-$ and $u-$channel diagrams. We then
describe the different processes by listing the contributing types of
diagrams, and specifying the couplings in terms of elementary vertex factors.
The latter are listed in an Appendix.

\subsection{General Formula}

In this subsection we present the squared spin and color averaged matrix
elements for squark pair production. We do not consider gluon fusion. Since we
only consider production of first and second generation squarks, we ignore all
quark mass effects, including mixing between $SU(2)$ doublet and singlet
squarks.

We begin by specifying functions that describe the contributions from various
kinds of matrix elements. $\Phi$ and $\chi$ describe squared $t-$ and
$u-$channel (gaugino exchange) diagrams\footnote{This includes products of two
  different $t-$ or $u-$channel diagrams.}, while $\Psi$ describes the
interference between a $t-$ channel and a $u-$channel diagram. Similarly,
$s-$channel (gauge boson exchange) and the interference between $s-$ and
$t-$channel diagrams are described by $\Upsilon$ and $\Omega$, respectively:
\begin{eqnarray}
\Phi(\tilde q_{i\alpha},\tilde q_{j\beta}^\prime,a) &=& \frac{1}{4} \sum_{l,k}
c_a(l,k) \frac{1}{\hat t-m_l^2} \frac{1}{\hat t-m_k^2} \Big\{ A(l,k,\tilde
q_{i\alpha},\tilde q_{j\beta}^\prime) 
\nonumber\\
&& \hspace*{4mm} \times (\hat t\hat u-m_{\tilde q_{i\alpha}}^2 m_{\tilde
  q_{j\beta}^\prime}^2) + B(l,k,\tilde q_{i\alpha},\tilde q_{j\beta}^\prime)
m_l m_k \hat s\Big\}\, , 
\nonumber\\
\chi(\tilde q_{i\alpha},\tilde q_{j\beta}^\prime,a) &=& \frac{1}{4} \sum_{l,k}
  c_a(l,k) \frac{1}{\hat u-M_l^2} \frac{1}{\hat u-M_k^2} \Big\{ C(l,k,\tilde
  q_{i\alpha},\tilde q_{j\beta}^\prime) 
\nonumber\\
&& \hspace*{4mm} \times (\hat t\hat u-m_{\tilde q_{i\alpha}}^2 m_{\tilde
  q_{j\beta}^\prime}^2)+ D(l,k,\tilde q_{i\alpha},\tilde q_{j\beta}^\prime)
  M_l M_k \hat s\Big\}\, , 
\nonumber\\
\Psi(\tilde q_{i\alpha},\tilde q_{j\beta}^\prime,a) &=& \frac{1}{4} \sum_{l,k}
  c_a(l,k)\frac{1}{\hat t-m_l^2}\frac{1}{\hat u-M_k^2} F(l,k,\tilde
  q_{i\alpha},\tilde q_{j\beta}^\prime) m_l M_k \hat s \, ,
\nonumber
\end{eqnarray}
\begin{eqnarray}
\Upsilon(q_g,q_h^\prime,\tilde q_{i\alpha},\tilde q_{j\beta}^\prime,a) &=&
\frac{1}{4} \sum_{l,k} c_a(l,k) \frac{1}{\hat s-M_l^2} \frac{1}{\hat s-M_k^2}
G(l,k,q_g,q_h^\prime,\tilde q_{i\alpha},\tilde q_{j\beta}^\prime) 
\nonumber\\
&& \hspace*{4mm} \times \Big\{ (m_{\tilde q_{j\beta}^\prime}^2 - m_{\tilde
  q_{i\alpha}}^2 + \hat t - \hat u) ( m_{\tilde q_{j\beta}^\prime}^2 -
m_{\tilde q_{i\alpha}}^2 + \hat u - \hat t)
\nonumber\\
&& \hspace*{6mm} - \hat s ( 2m_{\tilde q_{i\alpha}}^2 + 2m_{\tilde
  q_{j\beta}^\prime}^2 - \hat s)\Big\}\, ,
\nonumber\\
\Omega(q_g,q_h^\prime,\tilde q_{i\alpha},\tilde q_{j\beta}^\prime,a) &=&
 -\frac{1}{4} \sum_{l,k} c_a(l,k) \frac{1}{\hat s-M_l^2} \frac{1}{\hat
 t-m_k^2} H(l,k,q_g,q_h^\prime,\tilde q_{i\alpha},\tilde q_{j\beta}^\prime)
\nonumber\\
&& \hspace*{4mm} \times \Big\{ (m_{\tilde q_{j\beta}^\prime}^2 - \hat t)
(m_{\tilde  q_{i\alpha}}^2 - m_{\tilde q_{j\beta}^\prime}^2 + \hat u -\hat t)
\\
&& \hspace*{6mm}- \hat s (\hat s - 3m_{\tilde q_{j\beta}^\prime}^2 - m_{\tilde
  q_{i\alpha}}^2) + (m_{\tilde q_{j\beta}^\prime}^2 - \hat u) (m_{\tilde
  q_{i\alpha}}^2 - m_{\tilde q_{j\beta}^\prime}^2 + \hat t- \hat u)\Big\} \, . 
\nonumber  \label{master}
\end{eqnarray}
$\hat t$, $\hat u$ and $\hat s$ denote the partonic Mandelstam variables.
$m_{l,k}$, $M_{l,k}$ and $m_{\tilde q_{i\alpha, j\beta}}$ are the masses of
the propagating particles and the final states squarks, respectively; we use
capital letters for the masses of particles exchanged in the $u-$ or
$s-$channel, and lower case letters for masses in $t-$channel propagators. The
electrically neutral gauge bosons, all of which can contribute to the same
processes, are labeled through the indices $l,k=1,2,3$ for $\gamma$, $Z$ and
gluon, respectively; $W$-boson exchange can only occur in different reactions
than the exchange of the neutral gauge bosons. Similarly, the four neutralinos
and the gluino, which can contribute to the same process, are labeled by
$l,k=1,2,3,4,5$; alternatively, the two charginos are represented by
$l,k=1,2$. The flavour of the quarks and squarks is given by $q,q^\prime=u,d$.
$g,h,i, j=1,2$ are generation indices.  $\alpha,\beta=1,2$ label $SU(2)$
doublet ($L-$type) and singlet ($R-$type) squarks, respectively.  $c_a(l,k)$
are the colour factors for the different contributions, where $a$ labels the
various exchange topologies; note that unlike $l$ and $k$, $a$ is not summed.
Finally, the functions $A, B, C, D, F, G$ and $H$ are products of the various
coupling constants appearing in the matrix elements for the different
processes. Their general structure is given by
\begin{eqnarray} \label{coupl}
A(l,k,\tilde q_{i\alpha},\tilde q_{j\beta}^\prime) &=& a(l,\tilde
q_{i\alpha})a(k,\tilde q_{i\alpha})b^\prime(l,\tilde 
q_{j\beta}^\prime)b^\prime(k,\tilde q_{j\beta}^\prime)
\nonumber\\
&+& b(l,\tilde q_{i\alpha})b(k,\tilde q_{i\alpha})a^\prime(l,\tilde
q_{j\beta}^\prime)a^\prime(k,\tilde q_{j\beta}^\prime)\, ,
\nonumber\\
B(l,k,\tilde q_{i\alpha},\tilde q_{j\beta}^\prime) &=& a(l,\tilde
q_{i\alpha}) a(k,\tilde q_{i\alpha}) a^\prime(l,\tilde q_{j\beta}^\prime)
a^\prime(k,\tilde q_{j\beta}^\prime)
\nonumber\\ 
&+& b(l,\tilde q_{i\alpha}) b(k,\tilde q_{i\alpha}) b^\prime(l,\tilde
q_{j\beta}^\prime) b^\prime(k,\tilde q_{j\beta}^\prime)\,,
\nonumber\\
C(l,k,\tilde q_{i\alpha},\tilde q_{j\beta}^\prime) &=& c(l,\tilde
q_{j\beta}) c(k,\tilde q_{j\beta}) d^\prime(l,\tilde q_{i\alpha}^\prime)
d^\prime(k,\tilde q_{i\alpha}^\prime)
\nonumber\\
&+& d(l,\tilde q_{j\beta}) d(k,\tilde q_{j\beta}) c^\prime(l,\tilde
q_{i\alpha}^\prime) c^\prime(k,\tilde q_{i\alpha}^\prime)\,,
\nonumber\\
D(l,k,\tilde q_{i\alpha},\tilde q_{j\beta}^\prime) &=& c(l,\tilde
q_{j\beta})c(k,\tilde q_{j\beta}) c^\prime(l,\tilde q_{i\alpha}^\prime)
c^\prime(k,\tilde q_{i\alpha}^\prime)
\nonumber\\ 
&+& d(l,\tilde q_{j\beta}) d(k,\tilde q_{j\beta}) d^\prime(l,\tilde
q_{i\alpha}^\prime) d^\prime(k,\tilde q_{i\alpha}^\prime)\,,
\nonumber\\
F(l,k,\tilde q_{i\alpha},\tilde q_{j\beta}^\prime) &=& a(l,\tilde
q_{i\alpha}) c(k,\tilde q_{j\beta}) a^\prime(l,\tilde q_{j\beta}^\prime)
c^\prime(k,\tilde q_{i\alpha}^\prime)
\nonumber\\ 
&+& b(l,\tilde q_{i\alpha}) d(k,\tilde q_{j\beta}) b^\prime(l,\tilde
q_{j\beta}^\prime) d^\prime(k,\tilde q_{i\alpha}^\prime)\,,
\nonumber\\
G(l,k,q_g,q_h^\prime,\tilde q_{i\alpha},\tilde q_{j\beta}^\prime) &=&
c(l,\tilde q_{i\alpha},\tilde q_{j\beta}^\prime) c(k,\tilde q_{i\alpha},\tilde
q_{j\beta}^\prime) \nonumber \\
&\times& \big\{ e(l,q_g,q_h^\prime) e(k,q_g,q_h^\prime) +
f(l,q_g,q_h^\prime) f(k,q_g,q_h^\prime)\big\}\,,
\nonumber\\
H(l,k,q_g,q_h^\prime,\tilde q_{i\alpha},\tilde q_{j\beta}^\prime) &=&
c(l,\tilde q_{i\alpha},\tilde q_{j\beta}^\prime) \big\{ e(l,q_g,q_h^\prime)
b^\prime(k,\tilde q_{j\beta}^\prime) a(k,\tilde q_{i\alpha})
\nonumber\\ 
&&\hspace*{2cm} + f(l,q_g,q_h^\prime) a^\prime(k,\tilde q_{j\beta}^\prime)
b(k,\tilde q_{i\alpha}) \big\}\,.
\end{eqnarray}
Here $l$ and $k$ again label the exchanged (s)particles. $a\,, b\,, c\,, d\,,
a^\prime\,, b^\prime\,, c^\prime$ and $d^\prime$ denote couplings of the
relevant gaugino--quark--squark vertices; $a\,, c\,, a^\prime$ and $c^\prime$
denote left--handed couplings, i.e. the corresponding vertex factors are
multiplied with the left--chiral projector $P_L = (1-\gamma_5)/2$, while $b\,,
d\,, b^\prime$ and $d^\prime$ denote right--handed couplings. Similarly, $e$
and $f$ are left-- and right--handed gauge boson--quark--anti-quark couplings,
respectively, and $c$ is a gauge boson--squark--anti-squark coupling.

In the subsequent description of the contributing classes of subprocesses we
will specify the contributing diagrams, color factors, and couplings; explicit
expressions for the latter are given in the Appendix.

\subsection{$q q^\prime\rightarrow \tilde q\tilde q^\prime$}

We begin by discussing processes with two squarks or two anti--squarks in the
final state. There are no $s-$channel contributions to this class of
processes. We take all squark--quark--neutralino and squark--quark--gluino
couplings to be flavor--diagonal. Similarly, we ignore CKM mixing for charged
currents, i.e. we assume that quark--squark--chargino couplings only occur
within one generation. In the given class of reactions the flavor in the
initial and final state is therefore always the same.

\subsubsection{$u_i u_j\rightarrow \tilde u_{i\alpha}\tilde u_{j\beta}$}

\begin{figure}[h!]
\begin{center}
\hspace{-1.5cm}
\SetScale{0.5}
\begin{picture}(200,200)(0,0)
\Text(75,140)[lb]{$u_j$}
\Text(75,55)[]{{$ u_i$}}
\Text(120,100)[]{$\tilde g,{\tilde \chi}^0_m$}
\Text(125,140)[lb]{$\tilde{u}_{j\beta}$}
\Text(125,55)[lb]{${\tilde{u}}_{i\alpha}$}
\DashArrowLine(200,245)(238,327){2}
\DashArrowLine(200,155)(238,63){2}
\ArrowLine(162,327)(200,245)
\ArrowLine(162,63)(200,155)
\Line(200,155)(200,245)  
\Gluon(200,155)(200,245) {5}{4}
\Vertex(200,155){3.0}
\Vertex(200,245){3.0}
\end{picture}
\raisebox{3.5cm}{+}
\SetScale{0.5}
\begin{picture}(200,200)(0,0)
\Text(75,135)[lb]{{$u_i$}}
\Text(75,60)[lb]{{$u_i$}}
\Text(80,100)[]{{$\tilde g,{\tilde \chi}^0_m$}}
\Text(125,130)[lb]{{$\tilde{u}_{i\beta}$}}
\Text(125,65)[lb]{{$\tilde{u}_{i\alpha}$}}
\DashArrowLine(200,245)(238,163){2}
\DashArrowLine(200,155)(238,237){2}
\ArrowLine(111,261)(200,245)
\ArrowLine(111,139)(200,155)
\Line(200,155)(200,245) 
\Gluon(200,155)(200,245){5}{4}
\Vertex(200,155){3.0}
\Vertex(200,245){3.0}
\end{picture}
\caption{Feynman diagrams contributing to $u_i u_j \rightarrow \tilde
  u_{i\alpha} \tilde u_{j\beta}$. Here $i$ and $j$ are flavor indices, while
  $\alpha$ and $\beta$ label the `chirality' of the squarks, with $1\ (2)$
  standing for $L-$type ($R-$type) squarks. The index $m \in\{1,2,3,4\}$
  labels the exchanged neutralino. The second, $u-$channel, diagram only
  exists for $i = j$.
 \label{figF1}}
\end{center}
\end{figure}
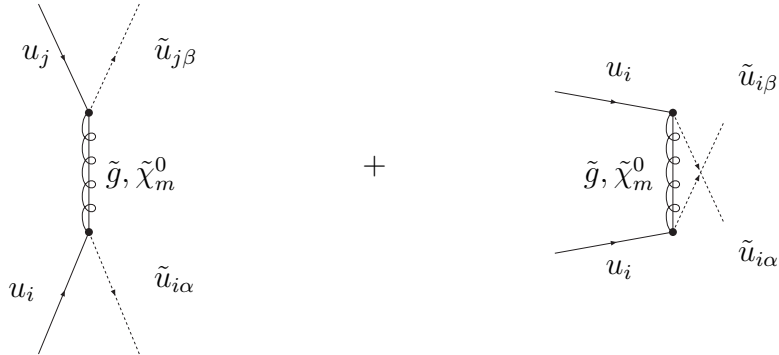

These processes proceed through the exchange of a neutralino or gluino in the
$t-$ or $u-$channel, as shown in Fig.~\ref{figF1}. 

In the notation of Eq.(\ref{master}) the squared spin-- and color--averaged
matrix element is given by
\begin{equation} \label{Muu}
\overline{|M|}^2 = \Phi(\tilde u_{i\alpha},\tilde u_{j\beta},1) + 
\chi(\tilde u_{i\alpha},\tilde u_{i\beta},1) \delta_{ij} +
\Psi(\tilde u_{i\alpha},\tilde u_{i\beta},2)\delta_{ij}. 
\end{equation}
If the final state squarks are identical, a statistics factor of $\frac{1}{2}$
must be included.  The colour factors of the $t-$ and the $u-$channel are
given by $c_1(l,k)$, while the factors for the interference term are given by
$c_2(l,k)$. Explicitly,
\begin{eqnarray} \label{Cuu}
c_1(l,k)=\left(\begin{array}{ccccc}
1 & 1 & 1 & 1 & 0\\
1 & 1 & 1 & 1 & 0\\
1 & 1 & 1 & 1 & 0\\
1 & 1 & 1 & 1 & 0\\
0 & 0 & 0 & 0 & 2/9
\end{array}\right),\quad
c_2(l,k)=\left(\begin{array}{ccccc}
1 & 1 & 1 & 1 & 4/9\\
1 & 1 & 1 & 1 & 4/9\\
1 & 1 & 1 & 1 & 4/9\\
1 & 1 & 1 & 1 & 4/9\\
4/9 & 4/9 & 4/9 & 4/9 & -2/27
\end{array}\right)\,.
\end{eqnarray}

The relevant neutralino--squark--quark and gluino--squark--quark-couplings to
be inserted in Eqs.(\ref{coupl}) are
\begin{eqnarray} \label{coupuu}
a(l,\tilde u_{i\alpha}) &=& a_{\tilde\chi^0_l/\tilde g}(\tilde
u_{i\alpha}),\qquad  \,\, b(l,\tilde u_{i\alpha})\;\, =\;\, b_{\tilde\chi^0_l/\tilde
  g}(\tilde u_{i\alpha}) \,,
\nonumber\\ 
a^\prime(l,\tilde u_{j\beta})  &=& a_{\tilde\chi^0_l/\tilde g}(\tilde
u_{j\beta}),\qquad  b^\prime(l,\tilde u_{j\beta}) \;\,=\;\,b_{\tilde
  \chi^0_l/\tilde g}(\tilde u_{j\beta}) \,,
\nonumber\\ 
c(l,\tilde u_{i\beta})  &=& a_{\tilde \chi^0_l/\tilde g}(\tilde
u_{i\beta}),\qquad  \,\, d(l,\tilde u_{i\beta})\;\, =\;\, b_{\tilde \chi^0_l/\tilde
  g}(\tilde u_{i\beta})\,,
\nonumber\\ 
c^\prime(l,\tilde u_{i\alpha})&=& a_{\tilde \chi^0_l/\tilde g}(\tilde
u_{i\alpha}),\qquad d^\prime(l,\tilde u_{i\alpha})\;\,=\;\, b_{\tilde
  \chi^0_l/\tilde g}(\tilde u_{i\alpha})\, .
\end{eqnarray}
As indicated earlier, $l\in\{1,2,3,4\}$ refers to the $l-$th neutralino, while
$l=5$ refers to the gluino. Explicit expressions for the neutralino and gluino
couplings appearing in Eq.(\ref{coupuu}) can be found in the Appendix,
Eqs.(\ref{neutcoup}) and (\ref{gluinocoup}).

Eqs.(\ref{Muu}) and (\ref{Cuu}) also hold for the charge conjugate
process. However, we have to use the appropriate anti--(s)quark couplings in
Eqs.(\ref{coupl}), which are given by
\begin{equation} \label{swapuu}
a_{\tilde \chi^0_l/\tilde g}(\bar{\tilde u}_{i\alpha}) = \left[ b_{\tilde
  \chi^0_l/\tilde g}(\tilde u_{i\alpha})\right]^*,\quad
b_{\tilde \chi^0_l/\tilde g}(\bar{\tilde u}_{i\alpha}) = \left[ a_{\tilde
  \chi^0_l/\tilde g}(\tilde u_{i{\alpha}})\right]^*\,,
\end{equation}
where $\bar{\tilde q}$ denotes an anti--squark and the stars stand for complex
conjugation. Note that even in a CP--conserving scenario some neutralino
couplings have to be complex (more exactly, purely imaginary) if we insist on
using positive neutralino masses in all propagators \cite{gh}. If CP is
violated, all chargino, neutralino and gluino couplings may be complex.
Finally, recall that a right--handed anti--quark is an $SU(2)$ doublet; its
couplings are therefore related to those of left--handed quarks, and vice
versa.

\subsubsection{$d_i d_j\rightarrow \tilde d_{i\alpha} \tilde d_{j\beta}$}
The process $d_i d_j\rightarrow \tilde d_{i\alpha} \tilde d_{j\beta}$ and its
charge--conjugated process are given by Eqs.(\ref{Muu}) to (\ref{swapuu}),
with the obvious replacement $\tilde u \rightarrow \tilde d$ everywhere.

\subsubsection{$u_i d_j\rightarrow \tilde u_{i\alpha} \tilde d_{j\beta}$}

\begin{figure}[h!]
\begin{center}
\hspace{-1.5cm}
\SetScale{0.5}
\begin{picture}(200,200)(0,0)
\Text(75,140)[lb]{$d_j$}
\Text(75,55)[]{{$ u_i$}}
\Text(120,100)[]{$\tilde g,{\tilde \chi}^0_m$}
\Text(125,140)[lb]{$\tilde{d}_{j\beta}$}
\Text(125,55)[lb]{${\tilde{u}}_{i\alpha}$}
\DashArrowLine(200,245)(238,327){2}
\DashArrowLine(200,155)(238,63){2}
\ArrowLine(162,327)(200,245)
\ArrowLine(162,63)(200,155)
\Line(200,155)(200,245)  
\Gluon(200,155)(200,245) {5}{4}
\Vertex(200,155){3.0}
\Vertex(200,245){3.0}
\end{picture}
\raisebox{3.5cm}{+}
%
\SetScale{0.5}
\begin{picture}(200,200)(0,0)
\Text(75,135)[lb]{{$d_i$}}
\Text(75,60)[lb]{{$u_i$}}
\Text(90,100)[]{{$\tilde{\chi}^+_n $}}
\Text(125,130)[lb]{{$\tilde{d}_{i\beta}$}}
\Text(125,65)[lb]{{$\tilde{u}_{i\alpha}$}}
\DashArrowLine(200,245)(238,163){2}
\DashArrowLine(200,155)(238,237){2}
\ArrowLine(111,261)(200,245)
\ArrowLine(111,139)(200,155)
\Line(200,155)(200,245) 
\Photon(200,155)(200,245){4}{4.5}
\Vertex(200,155){3.0}
\Vertex(200,245){3.0}
\end{picture}
%
\caption{Feynman diagrams contributing to $u_i d_j \rightarrow \tilde
  u_{i\alpha} \tilde d_{j\beta}$. The notation in the $t-$channel diagram is
  as in Fig.~\ref{figF1}. In the second, $u-$channel, diagram, which only
  exists of $i=j$, the chargino index $n$ runs from 1 to 2.
 \label{figF2}}
\end{center}
\end{figure}
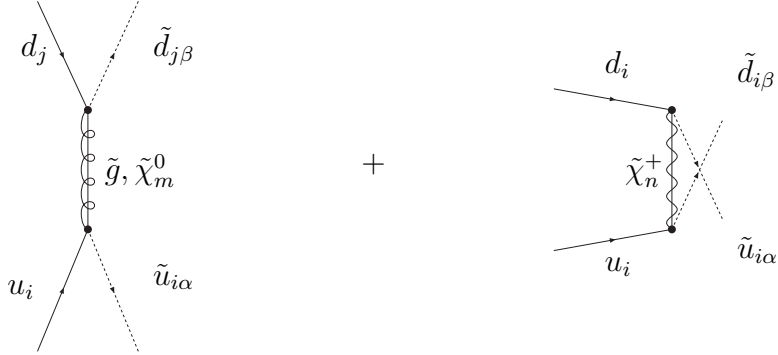

This process receives contributions from the $t-$channel exchange of a
neutralino or gluino; if both (s)quarks are from the same generation, $i=j$,
there is also a $u-$channel chargino exchange contribution. The corresponding
Feynman diagrams are shown in Fig.~\ref{figF2}.

The squared spin-- and color--averaged matrix element is given by
\begin{equation} \label{Mud}
\overline{|M|}^2 = \Phi(\tilde u_{i\alpha},\tilde d_{j\beta},1)
+ \chi(\tilde u_{i\alpha}, \tilde d_{i\beta},2)\delta_{ij}
+ \Psi(\tilde u_{i\alpha},\tilde d_{i\beta},3)\delta_{ij} \,. 
\end{equation}
The color factors for the squared $t-$channel neutralino and gluino
contributions, squared $u-$channel chargino contributions and of the
interference terms are given by
\begin{equation} \label{Cud}
c_1(l,k) = \left(\begin{array}{ccccc}
1 & 1 & 1 & 1 & 0\\
1 & 1 & 1 & 1 & 0\\
1 & 1 & 1 & 1 & 0\\
1 & 1 & 1 & 1 & 0\\
0 & 0 & 0 & 0 & 2/9
\end{array}\right),\quad
c_2(l,k)=\left(\begin{array}{cc}
1 & 1\\
1 & 1 
\end{array}\right),\quad
c_3(l,k)=\left(\begin{array}{cc}
1 & 1\\
1 & 1\\
1 & 1\\
1 & 1\\
4/9 & 4/9
\end{array}\right)\, .
\end{equation}
In the squared $t-$channel diagrams the indices $l,k$ labeling the exchanged
particles run from 1 to 5 for the four neutralinos and gluino, whereas in the
squared $u-$channel chargino--exchange contribution the indices run from 1 to
2. In the interference contribution the index $l$ labeling the particle
exchanged in the $t-$channel again runs from 1 to 5, while $k$ takes the
values 1 or 2.  The couplings to be inserted in Eqs.(\ref{coupl}) are given by
\begin{eqnarray} \label{coupud}
\begin{array} {rclrcl}
a(l,\tilde u_{i\alpha})& =& a_{\tilde \chi^0_l/\tilde g}(\tilde
u_{i\alpha}),& \quad b(l,\tilde u_{i\alpha})& =& b_{\tilde\chi^0_l/\tilde
  g}(\tilde u_{i\alpha})\,, \\ 
a^\prime(l,\tilde d_{j\beta})& =& a_{\tilde \chi^0_l/\tilde g}(\tilde
d_{j\beta}),&\quad b^\prime(l,\tilde d_{j\beta})& =& b_{\tilde \chi^0_l/\tilde
  g}(l,\tilde d_{j\beta})\,,
\\ 
c(l,\tilde d_{i\beta}) &= &a_{\tilde \chi^+_l}(\tilde d_{i\beta}), &\quad 
d(l,\tilde d_{i\beta}) & = & b_{\tilde \chi^+_l}(\tilde d_{i\beta})\,,
\\ 
c^\prime (l,\tilde u_{i\alpha}) & = & a_{\tilde \chi^+_l}(\tilde
u_{i\alpha}), &\quad d^\prime(l,\tilde u_{i\alpha}) & = & b_{\tilde \chi^+_l}(\tilde
u_{i\alpha}) 
\end{array}
\,. 
\end{eqnarray}
The explicit expressions for the couplings appearing in Eqs.(\ref{coupud}) can
be found in Eqs.(\ref{neutcoup}), (\ref{gluinocoup}) and (\ref{charcoup}) in
the Appendix.

The cross section for the charge--conjugated process can again be obtained by
using the appropriate couplings for anti--(s)quarks in Eqs.(\ref{coupl}):
\begin{eqnarray} \label{swapud}
\begin{array}{rclrcl}
a_{\tilde \chi^0_l/\tilde g/\tilde \chi^+_l}(\bar{\tilde d}_{i\alpha}) &=&
  \left[ b_{\tilde \chi^0_l/ \tilde g/\tilde \chi^+_l}(\tilde d_{i\alpha})
  \right]^*\,, &\quad
b_{\tilde \chi^0_l/\tilde g/\tilde \chi^+_l}(\bar{\tilde d}_{i\alpha}) &=& 
\left[ a_{\tilde \chi^0_l/ \tilde g/\tilde\chi^+_l}(\tilde d_{i{\alpha}})
  \right]^* \, , 
\\
a_{\tilde \chi^0_l/\tilde g/\tilde\chi^+_l}(\bar{\tilde u}_{i\alpha}) &=& \left[
  b_{\tilde \chi^0_l,/ \tilde g/\tilde \chi^+_l}(\tilde u_{i\alpha}) \right]^*
  \, , &\quad 
b_{\tilde \chi^0_l/\tilde g/\tilde\chi^+_l}(\bar{\tilde u}_{i\alpha}) &=& \left[
  a_{\tilde \chi^0_l/ \tilde g/\tilde \chi^+_l}(\tilde u_{i\alpha}) \right]^*
\end{array}  \, .
\end{eqnarray}
\subsubsection{$u_i d_j\rightarrow \tilde d_{i\alpha}\tilde u_{j\beta},\quad i
  \neq j$} 

\begin{figure}[h!]
\begin{center}
\hspace{-1.5cm}
\SetScale{0.5}
\begin{picture}(200,200)(0,0)
\Text(75,140)[lb]{$d_j$}
\Text(75,55)[]{{$ u_i$}}
\Text(120,100)[]{$\tilde {\tilde \chi}^+_n$}
\Text(125,140)[lb]{$\tilde{u}_{j\beta}$}
\Text(125,55)[lb]{${\tilde{d}}_{i\alpha}$}
\DashArrowLine(200,245)(238,327){2}
\DashArrowLine(200,155)(238,63){2}
\ArrowLine(162,327)(200,245)
\ArrowLine(162,63)(200,155)
\Line(200,155)(200,245)  
\Gluon(200,155)(200,245) {5}{4}
\Vertex(200,155){3.0}
\Vertex(200,245){3.0}
\end{picture}
\caption{Feynman diagram contributing to $u_i d_j \rightarrow \tilde
  u_{j\alpha} \tilde d_{i\beta}$ with $i \neq j$. The notation is
  as in the chargino exchange diagram of Fig.~\ref{figF2}.
 \label{figF3}}
\end{center}
\end{figure}
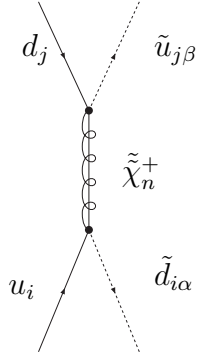

This process differs from the one considered in the previous subsection only
if the two (s)quarks are from {\em different} generation, with the $d-$type
squark in the final state being from the same generation as the initial
$u-$type quark. In this case only the chargino exchange diagram shown in
Fig.~\ref{figF3} contributes. We label the momenta such that this is a
$t-$channel diagram.

The squared spin-- and color--averaged matrix element is given by
\begin{equation} \label{Mud1}
\overline{|M|}^2=\Phi(\tilde d_{i\alpha},\tilde u_{j\beta},1)\,.
\end{equation}
The colour factor is trivial,
\begin{equation} \label{Cud1}
c_1=\left(\begin{array}{cc}
1 & 1\\
1 & 1
\end{array}\right)\,.
\end{equation}
The couplings to be inserted in Eqs.(\ref{coupl}) are,
\begin{eqnarray} \label{coupud1}
\begin{array}{rclrcl}
a(l,\tilde d_{i\alpha}) &=& a_{\tilde \chi^+_l}(\tilde d_{i\alpha})\,,&\quad
b(l,\tilde d_{i\alpha}) &=& b_{\tilde \chi^+_l}(\tilde d_{i\alpha})\,,
\\ 
a^\prime (l,\tilde u_{j\beta}) &=& a_{\tilde \chi^+_l}(\tilde
u_{j\beta}),&\quad b^\prime(l,\tilde u_{j\beta}) &=& b_{\tilde \chi^+_l}(\tilde
u_{j\beta}) \end{array}\,.
\end{eqnarray}
The cross section for the charge conjugated process can be obtained by using
the appropriate anti--(s)quark couplings, which have already been given in
Eqs.(\ref{swapud}).

\subsection{$q \bar q^{\prime} \rightarrow \tilde q \bar{\tilde q}$}

We now turn to processes with one squark and one anti--squark in the final
state. Since now the initial and final state have vanishing baryon charge,
$s-$channel diagrams may contribute.

\subsubsection{$u_i \bar u_j\rightarrow \tilde u_{i\alpha} \bar{\tilde
    u}_{j\beta} $}

\begin{figure}[h!]
\begin{center}
\hspace{-1.5cm}
\SetScale{0.5}
\begin{picture}(200,200)(0,0)
\Text(75,140)[lb]{{$\bar u_j$}}
\Text(75,55)[lb]{{$ u_i$}}
\Text(120,100)[]{{$\tilde g,\tilde{\chi}^0_m $}}
\Text(125,140)[lb]{{$\bar{\tilde u}_{j\beta}$}}
\Text(125,55)[lb]{{${\tilde{u}}_{i\alpha}$}}
\DashArrowLine(238,327)(200,245){2}
\DashArrowLine(200,155)(238,63){2}
\ArrowLine(200,245)(162,327)
\ArrowLine(162,63)(200,155)
\Line(200,155)(200,245)  
\Gluon(200,155)(200,245) {5}{4}
\Vertex(200,155){3.0}
\Vertex(200,245){3.0}
\end{picture}
\raisebox{3.5cm}{+}
\SetScale{0.5}
\begin{picture}(200,200)(0,0)
\Text(50,125)[lb]{{$\bar u_i$}}
\Text(50,65)[lb]{{$ u_i$}}
\Text(100,90)[]{{$\gamma,Z,g$}}
\Text(150,125)[lb]{{$\bar{\tilde u}_{i\alpha}$}}
\Text(150,65)[lb]{{${\tilde{u}}_{i\alpha}$}}
\DashArrowLine(333,245)(245,200){2}
\DashArrowLine(245,200)(333,155){2}
\ArrowLine(155,200)(67,245)
\ArrowLine(67,155)(155,200)
\Photon(155,200)(245,200) {4}{4.5}
\Vertex(155,200){3.0}
\Vertex(245,200){3.0}
\end{picture}
\caption{Feynman diagrams contributing to $u_i \bar u_j\rightarrow \tilde
  u_{i\alpha} \bar{\tilde u}_{j\beta}$. The notation for the $t-$channel
  diagram is as in Fig.\ref{figF1}. The gauge boson exchanged in the second,
  $s-$channel, diagram, which only exists if $i=j$, can be a gluon, a photon
  or a $Z$ boson. 
 \label{figF4}}
\end{center}
\end{figure}
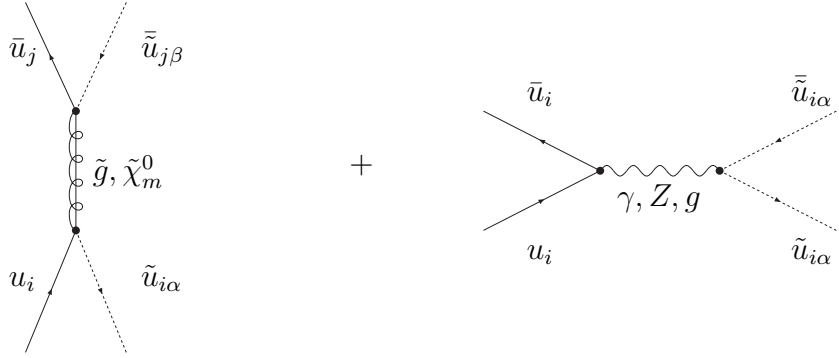

This process receives contributions from the exchange of a gluino or
neutralino in the $t-$channel; if $i = j$, there are also $s-$channel gluon,
photon and $Z$ exchange contributions. The corresponding Feynman diagrams are
shown in Fig.~\ref{figF4}. 

The squared spin-- and color--averaged matrix element is given by
\begin{equation} \label{Muubar}
\overline{|M|}^2 = \Phi(\tilde u_{i\alpha},\bar{\tilde u}_{j\beta},1)
+ \Upsilon(u_i,\bar u_i,\tilde u_{i\alpha},\bar{\tilde u}_{i\alpha},2) 
\delta_{ij} \delta_{\alpha\beta} 
+ \Omega(u_i,\bar u_i,\tilde u_{i\alpha},\bar{\tilde u}_{i\alpha},3)
\delta_{ij} \delta_{\alpha\beta}\, .
\end{equation}
The color factors of the pure $t-$channel neutralino and gluino contributions,
the $s-$channel $\gamma, Z$, gluon contributions and the interference terms
are given by
\begin{eqnarray}  \label{Cuubar}
c_1(l,k)&=&\left(\begin{array}{cccccc}
1 & 1 & 1 & 1 & 0\\
1 & 1 & 1 & 1 & 0\\
1 & 1 & 1 & 1 & 0\\
1 & 1 & 1 & 1 & 0\\
0 & 0 & 0 & 0 & 2/9
\end{array}\right),\qquad
c_2(l,k)\;=\;\left(\begin{array}{ccc}
1 & 1 & 0 \\
1 & 1 & 0 \\
0 & 0 & 2/9\\
\end{array}\right),\nonumber\\[1.5mm]
c_3(l,k)&=&\left(\begin{array}{ccccc}
1 & 1 & 1 & 1 & 4/9\\
1 & 1 & 1 & 1 & 4/9\\
4/9 & 4/9 & 4/9 & 4/9 & -2/27\\
\end{array}\right)\,.
\end{eqnarray}
The couplings to be inserted in Eqs.(\ref{coupl}) are given by:
\begin{eqnarray} \label{coupuubar}
\begin{array}{rclrcl}
a(l,\tilde u_{i\alpha}) &=& a_{\tilde \chi^0_l/\tilde g}(\tilde u_{i\alpha})
\,, &\quad  
b(l,\tilde u_{i\alpha})& =& b_{\tilde \chi^0_l/ \tilde g}(\tilde u_{i\alpha})\,,
\\ 
a^\prime(l,\bar{\tilde u}_{j\beta})& =& \left[ b_{\tilde \chi^0_l/\tilde
    g}(\tilde u_{j\beta}) \right]^* \,,& \quad 
b^\prime(l,\bar{\tilde u}_{j\beta})& =& \left[ a_{\tilde \chi^0_l/\tilde
    g}(\tilde u_{j\beta}) \right]^* \, , 
\\[1.5mm]
e(l,u_i,\bar u_i)& =& e_{\gamma/Z/g}(u_i,\bar u_i)\,,&\quad f(l,u_i,\bar
u_i)& =& q_{\gamma/Z/g}(u_i,\bar u_i)\,,
\\ [1.5mm]
c(l,\tilde u_{i\alpha},\bar{\tilde u}_{i\alpha}) &=& c_{\gamma/Z/g}(\tilde
u_{i\alpha},\bar{\tilde u}_{i\alpha})\,.&&&
\end{array}
\end{eqnarray}
Recall that in $s-$channel diagrams $l=1$ stands for a photon, $l=2$ for a
$Z-$boson, and $l=3$ for a gluon. The explicit expressions for the couplings
of these gauge bosons can be found in Eqs.(\ref{coupef}) and (\ref{coupc}) in
the Appendix.

\subsubsection{$u_i \bar u_i\rightarrow \tilde q_{j\alpha} \bar{\tilde
    q}_{j\alpha} ,\quad  i \ne j$}

Since the flavor in the initial and final state is different, only the
$s-$channel diagrams of Fig.\ref{figF4} contribute. 
The squared spin-- and color--averaged matrix element is thus simply given by
\begin{equation} \label{Muubar1}
\overline{|M|}^2 = \Upsilon(u_i,\bar u_i,\tilde q_{i\alpha}, \bar{\tilde
q}_{i\alpha},1)\,.
\end{equation}
The colour factors are
\begin{equation} \label{Cuubar1}
c_1(l,k)=\left(\begin{array}{ccc}
1 & 1 & 0 \\
1 & 1 & 0 \\
0 & 0 & 2/9\\
\end{array}\right)\,.\\
\end{equation}
The couplings to be inserted in Eqs.(\ref{coupl}) can be read off from
Eqs.(\ref{coupuubar}).

\subsubsection{$u_i \bar u_j\rightarrow \tilde d_{i\alpha} \bar {\tilde
    d}_{j\beta}$} 

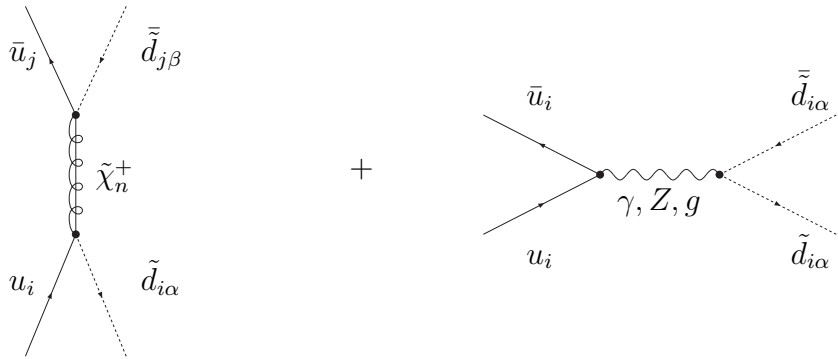
\begin{figure}[h!]
\begin{center}
\hspace{-1.5cm}
\begin{picture}(200,200)(0,0)
\SetScale{0.5}
\Text(75,140)[lb]{{$\bar u_j$}}
\Text(75,55)[lb]{{$ u_i$}}
\Text(115,100)[]{{$\tilde{\chi}^+_n $}}
\Text(125,140)[lb]{{$\bar{\tilde{d}}_{j\beta}$}}
\Text(125,55)[lb]{{${\tilde{d}}_{i\alpha}$}}
\DashArrowLine(238,327)(200,245){2}
\DashArrowLine(200,155)(238,63){2}
\ArrowLine(200,245)(162,327)
\ArrowLine(162,63)(200,155)
\Line(200,155)(200,245)  
\Gluon(200,155)(200,245) {5}{4}
\Vertex(200,155){3.0}
\Vertex(200,245){3.0}
\end{picture}
\raisebox{3.5cm}{+}
\SetScale{0.5}
\begin{picture}(200,200)(0,0)
\Text(50,125)[lb]{{$\bar u_i$}}
\Text(50,65)[lb]{{$ u_i$}}
\Text(100,90)[]{{$\gamma,Z,g$}}
\Text(150,125)[lb]{{$\bar{\tilde d}_{i\alpha}$}}
\Text(150,65)[lb]{{${\tilde{d}}_{i\alpha}$}}
\DashArrowLine(333,245)(245,200){2}
\DashArrowLine(245,200)(333,155){2}
\ArrowLine(155,200)(67,245)
\ArrowLine(67,155)(155,200)
\Photon(155,200)(245,200) {4}{4.5}
\Vertex(155,200){3.0}
\Vertex(245,200){3.0}
\end{picture}
\caption{Feynman diagrams contributing to $u_i \bar u_j\rightarrow \tilde
  d_{i\alpha} \bar {\tilde d}_{j\beta}$. The notation for the $t-$channel
  diagram is as in Fig.\ref{figF3}. The notation for the second,
  $s-$channel, diagram, which only exists if $i=j$, is as in Fig.~\ref{figF4}.
 \label{figF5}}
\end{center}
\end{figure}

This process receives contributions from chargino exchange in the $t-$channel;
if $i = j$, there are also $s-$channel contributions with gluon, photon and
$Z-$exchange. The corresponding Feynman diagrams are shown in
Fig.~\ref{figF5}. 

The squared spin-- and color--averaged matrix element is given by
\begin{equation} \label{Muubar2}
\overline{|M|}^2=\Phi(\tilde d_{i\alpha}, \bar{ \tilde d}_{j\beta},1)
+ \Upsilon(u_i,\bar u_i,\tilde d_{i\alpha},\bar{\tilde d}_{i\alpha},2)
\delta_{ij} \delta_{\alpha\beta}
+\Omega(u_i,\bar u_i,\tilde d_{i\alpha},\bar{\tilde d}_{i\alpha},3)
\delta_{ij} \delta_{\alpha\beta}\,.
\end{equation}
The respective colour factors for the squared $t-$channel, squared $s-$channel
and the interference terms are given by 
\begin{equation} \label{Cuubar2}
c_1=\left(\begin{array}{cc}
1 & 1\\
1 & 1\\
\end{array}\right)\,,\quad 
c_2(l,k)=\left(\begin{array}{ccc}
1 & 1 & 0 \\
1 & 1 & 0 \\
0 & 0 & 2/9\\
\end{array}\right)\,,\quad
c_3(l,k)=\left(\begin{array}{cc}
1 & 1\\
1 & 1\\
4/9 & 4/9
\end{array}\right)\,.
\end{equation}
The couplings to be inserted in Eqs.(\ref{coupl}) are given by
\begin{eqnarray} \label{coupuubar2}
\begin{array}{rclrcl}
a(l,\tilde d_{i\alpha}) &=& a_{\tilde \chi^+_l}(\tilde d_{i\alpha})
\,,&\quad
a^\prime(l,\bar{\tilde d}_{j\beta})& =& \left[ b_{\chi^+_l}(\tilde
d_{j\beta}) \right]^*\,,
\\ [2.3mm] 
b(l,\tilde d_{i\alpha})& =& b_{\tilde \chi^+_l}(\tilde d_{i\alpha})
\,,&\quad
b^\prime(l,\bar{\tilde d}_{j\beta})& =& \left[a_{\chi^+_l}(\tilde
d_{j\beta}) \right]^* \, ,
\\ [1.9mm]
e(l,u_i,\bar u_i)& =& e_{\gamma/Z/g}(u_i,\bar u_i)\,,&\quad
f(l,u_i,\bar u_i)& =& q_{\gamma/Z/g}(u_i,\bar u_i)\,,
\\
c(l,\tilde d_{i\alpha},\bar{\tilde d}_{i\alpha})& =& c_{\gamma/Z/g}(\tilde
d_{i\alpha},\bar{\tilde d}_{i\alpha})\,.
\end{array}
\end{eqnarray}

\subsubsection{$d_i\bar d_j\rightarrow \tilde q \bar{\tilde q}$}

Each of the last three processes has an analogue where all $u-$type (s)quarks
are replaced by $d-$type (s)quarks and vice versa. The cross sections for
these reactions can be described by simply replacing $u \rightarrow d$ and $d
\rightarrow u$ everywhere.

\subsubsection{$d_i\bar u_j\rightarrow \tilde d_{i\alpha} \bar{\tilde
    u}_{j\beta}$} 

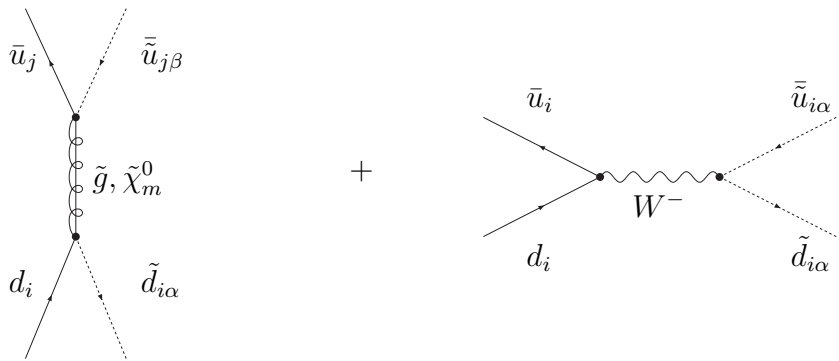
\begin{figure}[h!]
\begin{center}
\hspace{-1.5cm}
\begin{picture}(200,200)(0,0)
\SetScale{0.5}
\Text(75,140)[lb]{{$\bar u_j$}}
\Text(75,55)[lb]{{$ d_i$}}
\Text(120,100)[]{{$\tilde g,\tilde{\chi}^0_m $}}
\Text(125,140)[lb]{{$\bar{\tilde u}_{j\beta}$}}
\Text(125,55)[lb]{{${\tilde{d}}_{i\alpha}$}}
\DashArrowLine(238,327)(200,245){2}
\DashArrowLine(200,155)(238,63){2}
\ArrowLine(200,245)(162,327)
\ArrowLine(162,63)(200,155)
\Line(200,155)(200,245)  
\Gluon(200,155)(200,245) {5}{4}
\Vertex(200,155){3.0}
\Vertex(200,245){3.0}
\end{picture}
\raisebox{3.5cm}{+}
\SetScale{0.5}
\begin{picture}(200,200)(0,0)
\Text(50,125)[lb]{{$\bar u_i$}}
\Text(50,65)[lb]{{$ d_i$}}
\Text(100,90)[]{{$W^-$}}
\Text(150,125)[lb]{{$\bar{\tilde u}_{i\alpha}$}}
\Text(150,65)[lb]{{${\tilde{d}}_{i\alpha}$}}
\DashArrowLine(333,245)(245,200){2}
\DashArrowLine(245,200)(333,155){2}
\ArrowLine(155,200)(67,245)
\ArrowLine(67,155)(155,200)
\Photon(155,200)(245,200) {4}{4.5}
\Vertex(155,200){3.0}
\Vertex(245,200){3.0}
\end{picture}
\caption{Feynman diagrams contributing to $d_i\bar u_j\rightarrow \tilde
  d_{i\alpha} \bar{\tilde u}_{j\beta}$. The notation for the $t-$channel
  diagram is as in Fig.\ref{figF3}. The second, $s-$channel, diagram, which
  only exists if $i=j$, proceeds via the exchange of a charged $W$ boson.
 \label{figF6}}
\end{center}
\end{figure}

This process receives contributions from the exchange of a gluino or
neutralino in the $t-$channel; if $i=j$, there is also an $s-$channel $W$
exchange contribution. The corresponding Feynman diagrams are shown in
Fig.~\ref{figF6}.

The squared spin-- and color--averaged matrix element is given by
\begin{equation} \label{Mudbar}
\overline{|M|}^2 = \Phi(\tilde d_{i\alpha}, \bar{\tilde u}_{j\beta},1)
+ \Upsilon(d_i,\bar u_i,\tilde d_{i\alpha}, \bar{\tilde u}_{i\alpha},2)
\delta_{ij} \delta_{\alpha\beta}
+\Omega(d_i,\bar u_i,\tilde d_{i\alpha}, \bar{\tilde u}_{i\alpha},3)
\delta_{ij} \delta_{\alpha\beta}\,.
\end{equation}
The color factors for the pure $t-$channel, pure $s-$channel and interference
terms are
\begin{equation} \label{Cudbar}
c_1(l,k) = \left(\begin{array}{cccccc}
1 & 1 & 1 & 1 & 0\\
1 & 1 & 1 & 1 & 0\\
1 & 1 & 1 & 1 & 0\\
1 & 1 & 1 & 1 & 0\\
0 & 0 & 0 & 0 & 2/9
\end{array}\right)\, ,\quad
c_2 = 1\,,\quad 
c_3 = \left(\begin{array}{ccccc}1 & 1 & 1 & 1 & 4/9\end{array}\right) \,.
\end{equation}
The couplings to be inserted in Eqs.(\ref{coupl}) are
\begin{eqnarray} \label{coupudbar}
\begin{array}{rclrcl}
a(l,\tilde d_{i\alpha})& =& a_{\tilde \chi^0_l/\tilde g}(\tilde d_{i\alpha})
\,,& \quad
a^\prime(l,\bar{\tilde u}_{j\beta}) &=& \left[ b_{\tilde \chi^0_l/\tilde g}
  (\tilde u_{j\beta}) \right]^* \, ,
\\ [2.5mm]
b(l,\tilde d_{i\alpha}) &=& b_{\tilde \chi^0_l/\tilde g}(\tilde d_{i\alpha})
\,,&\quad 
b^\prime(l,\bar{\tilde u}_{j\beta}) &=& \left[a_{\tilde \chi^0_l/\tilde g}(\tilde
u_{j\beta}) \right]^* \,, \\[2.5mm] 
e(d_i,\bar u_i) &=& e_W(d_i,\bar u_i)\,,&\quad 
f(d_i,\bar u_i) &=& f_W(l,d_i,\bar u_i)\,,
\\  
c(l,\tilde d_{i\alpha},\bar{\tilde u}_{i\alpha})& =& c_W(\tilde d_{i\alpha},
\bar{\tilde u}_{i\alpha}) &&&
\end{array}
\,.
\end{eqnarray}
Explicit expressions for the couplings of the $W$ boson can be found in
Eqs.(\ref{coupW1}) and (\ref{coupW2}) in the Appendix.

Unlike for the processes discussed so far in this Subsection, charge
conjugation here leads to a physically different reaction. The cross section
for this process can be obtained from Eqs.(\ref{Mudbar})--(\ref{coupudbar}) by
replacing (s)quark couplings with anti--(s)quark couplings and vice versa. The
new couplings appearing in the $t-$channel diagrams can e.g. be read off from
Eqs.(\ref{swapud}), whereas the couplings in the $s-$channel diagram remain
unchanged.

\subsubsection{$d_i \bar u_i\rightarrow \tilde d_{j\alpha} \bar{\tilde
  u}_{j\beta}\,, \quad i\ne j$}

This process can only proceed through the exchange of a charged $W$ boson in
the $s-$channel. The corresponding Feynman diagram has already been shown in
Fig.~\ref{figF6}. The squared spin-- and color--averaged matrix element is
simply given by
\begin{equation} \label{Mudbar1}
\overline{|M|}^2 = \Upsilon(d_i,\bar u_i, \tilde d_{j\alpha}, \bar{\tilde
u}_{j\beta},1)\, .
\end{equation}
The color factor is trivial, $c_1 = 1$.  The couplings to be inserted in
Eqs.(\ref{coupl}) are
\begin{eqnarray} \label{coupludbar1}
e(l,d_i,\bar u_i) &=& e_W(d_i,\bar u_i)\,,\qquad 
f(l,d_i,\bar u_i)\;\, = \;\,f_W(l,d_i,\bar u_i)\,,
\nonumber\\ 
c(l,\tilde d_{j\alpha},\bar{\tilde u}_{j\beta}) &=& c_W(\tilde
d_{j\alpha},\bar{ \tilde u}_{j\beta})\, .
\end{eqnarray}
The squared matrix element for the charge conjugated process is identical.

\section{Numerical Results}

We are now ready to present numerical results. We focus on $pp$ collisions at
the LHC operating at $\sqrt{s} = 14$ TeV, since in most predictive models of
supersymmetry breaking \cite{susyrev} the existing bounds \cite{pdg} on the
masses of sleptons and charginos imply that first and second generation
squarks are too heavy to be produced at the Tevatron. Since our calculation is
leading order in QCD, we use CTEQ5L structure functions \cite{cteq}, as well
as the one--loop expression for the running QCD coupling $\alpha_s$, with five
active flavors and $\Lambda_{\rm QCD} = 142$ MeV. The electroweak gauge
couplings, as well as the relevant superparticle masses, are taken from the
output of the program package SoftSUSY \cite{softsusy}. The couplings are
$\overline {\rm MS}$ couplings at a scale near the squark masses, whereas the
squark and gluino masses are on--shell (pole) masses. We focus on squarks of
the first and second generation, where mixing between $SU(2)$ doublets and
singlets can be neglected. Third generation squarks are produced dominantly
through gluon fusion or pure $s-$channel diagrams; the EW contributions to
these cross sections will therefore be very small. Experimentally the
production of third generation squarks can be distinguished by detecting $b$
or $t$ quarks in the final state.

\begin{table}[t!] 
\begin{center}
\begin{tabular}{|c||r|r|r||c|c||c|c||c|c|} 
\hline
& & & & \multicolumn{2}{|c||}{QCD} &
\multicolumn{2}{|c||}{QCD $+$ EW} & \multicolumn{2}{|c|}{ratio} \\
Scenario & $m_0$ & $m_{1/2}$ & $m_{\tilde q}$ & Total & LL & Total & LL &
Total & LL \\ 
\hline \hline
SPS 1a & 100 & 250 & 560 & 12.11 & 3.09 & 12.55 & 3.50 & 1.036 & 1.133 \\
SPS 1b & 200 & 400 & 865 & 1.57 & 0.42 & 1.66 & 0.499 & 1.055 & 1.186 \\
SPS 2 & 1450 & 300 & 1590 & 0.0553 & 0.0132 & 0.0567 & 0.0144 & 1.025 & 1.091
\\ 
SPS 3 & 90 & 400 & 845 & 1.74 & 0.464 & 1.83 & 0.551 & 1.055 & 1.188 \\
SPS 4 & 400 & 300 & 760 & 3.10 & 0.813 & 3.22 & 0.927 & 1.040 & 1.141 \\
SPS 5 & 150 & 300 & 670 & 5.42 & 1.41 & 5.66 & 1.62 & 1.042 & 1.152 \\
\hline

\end{tabular}
\caption{Total cross sections at the LHC for combined first and second
  generation squark pair production from quark initial states in six mSUGRA
  benchmark scenarios. All masses are in GeV, $m_0$ and $m_{1/2}$ being the
  common soft breaking scalar and gaugino masses, respectively, at the scale
  of Grand Unification, and $m_{\tilde q}$ giving the average mass of first
  generation $SU(2)$ doublet squarks. All cross sections are in pb. The last
  two columns show the ratio (QCD $+$ EW) $/$ QCD. We show results for the sum
  over all squark pairs (``total''), as well as for the sum over all
  combinations of two $SU(2)$ doublet squarks (``LL''); in both cases we
  include squarks and anti--squarks. The cross sections have been calculated
  in leading order, using the CTEQ5L parton distribution functions
  \cite{cteq}.}
\end{center}
\end{table}

Table 1 shows results for the total squark pair production cross sections at
the LHC in six mSUGRA benchmark scenarios, taken from \cite{sps}. Here we sum
over all squarks and anti--squarks of the first and second generation; results
where both final state (anti--)squarks are $SU(2)$ doublets are shown
separately.  We only include contributions with (anti--)quarks in the initial
state, since the gluon fusion contribution obviously does not receive
electroweak contributions in leading order. Besides, the gluon fusion
contribution is subdominant in all cases; it increases the QCD contribution by
14 (6.7, 1.4)\% for SPS 1a (1b, 2).\footnote{The {\em inclusive} squark
  production cross section also receives large contributions from associate
  $\tilde q \tilde g$ production from pure QCD, as well as from $\tilde q
  \tilde \chi^{\pm,0}$ production at ${\cal O}(\alpha_s \alpha_w)$
  \cite{qcdlo}.} We take equal factorization and renormalization scales,
$\mu_F = \mu_R = m_{\tilde q} / 2$; this choice leads to quite small NLO
corrections to the pure QCD contribution \cite{qcdnlo}.  Increasing (reducing)
these scales will reduce (increase) the prediction for the pure QCD
contribution, thereby enhancing (diminishing) the relative importance of the
electroweak contributions.

Not surprisingly, the cross sections fall quickly with increasing squark mass.
The partonic cross sections scale like $m^{-2}_{\tilde q}$, if the ratios of
sparticle masses are kept fixed and the running of $\alpha_s$ is ignored. In
addition, the pdf factors decrease quickly with increasing squark mass. There
is also some dependence on the gluino mass ($\simeq 2.5 m_{1/2}$), which
appears in $t-$ and $u-$channel propagators. Varying the ratio $m_{\tilde g} /
m_{\tilde q}$ between 0.5 and 1.2, which is the range covered by the scenarios
of Table~1, leads to 15 to 20\% variation of the QCD prediction. We will
discuss the gaugino mass dependence in more detail later.

Since QCD contributions dominate even after inclusion of the electroweak
diagrams, the overall behavior of the total cross sections does not change
much. These contributions are clearly more important for the production of
$SU(2)$ doublet squarks than for the total cross section summed over all final
states. This is not surprising: the cross sections for all other combinations
of squarks only receive electroweak contributions due to hypercharge
interactions, and the squared $SU(2)$ gauge coupling exceeds the squared
$U(1)_Y$ coupling by a factor $\cot^2 \theta_W \simeq 3.3$.

However, the weakness of the $U(1)_Y$ coupling by itself is not sufficient to
explain the small size of electroweak contributions to final states involving
at least one $SU(2)$ singlet $\mbox{\rm (anti--)squark}$. For example, we can
infer from the first line of Table 1 that in scenario SPS 1a, electroweak
contributions increase the cross section for the production of two $L-$type
squarks by 0.41 pb, whereas they only contribute 0.03 pb to all other squark
pair production channels combined. We also note that the importance of the EW
contributions seems to depend much more strongly on the ratio $m_{1/2} / m_0$
than the QCD prediction does. Finally, the EW contributions evidently become
more important for heavier squarks if the ratio $m_0 / m_{1/2}$ remains
roughly the same.

In order to understand these features, in Table~2 we list results for all 24
different processes involving only (s)quarks and anti--(s)quarks from the first
generation. These results are for SPS 1a, but we find similar patterns in the
other scenarios.

These 24 processes can be grouped into three categories. The first seven
reactions show interference between $t-$ and $u-$channel diagrams, where in
all but the last case there are both strong and electroweak contributions from
both $t-$ and $u-$channel diagrams. The next class of seven processes allows
interference between $s-$ and $t-$channel diagrams. In the first four cases
there are both QCD and electroweak contributions to both the $t-$ and
$s-$channel, while in the last three cases only one QCD diagram contributes.
For the third class of ten processes, no interference between electroweak and
strong contributions is possible; two of these processes only proceed via
$s-$channel diagrams, whereas the remaining eight are pure $t-$channel
reactions.

Within processes proceeding through $t-$ or $u-$channel diagrams, it is
important to distinguish between reactions that require a helicity flip of the
exchanged fermion, and those that don't. Consider reactions with two squarks
with equal `chirality' in the final state, e.g. $\ul \ul$ production (process
no.~1 in Table~2). In this case both quarks in the initial state have to be
left--handed, since the quark--squark--gaugino gauge couplings couple $L-$type
squarks only to left--handed quarks. This evidently requires a helicity flip
of the exchanged gaugino, i.e. the corresponding amplitudes are proportional
to the mass of the exchanged gaugino. In models with gaugino mass unification
this further suppresses electroweak contributions in this important class of
processes, since the $SU(2)$ and $U(1)_Y$ gaugino masses are respectively
three and six times smaller than the gluino mass. On the other hand, the equal
helicities in the initial state imply that total angular momentum $J = 0$, so
that the produced squarks can be in an $S-$wave. The cross section therefore
receives only a single power of the threshold factor $\beta$ (which is the
squark center--of--mass [cms] velocity).

In contrast, the production of one $L-$type and one $R-$type squark,
e.g. $\ul \ur$ production (process no.~3 in Table~2), requires {\em opposite}
helicities of the two initial quarks. The exchanged fermion therefore does not
need to change helicity, i.e. these amplitudes remain finite even in the limit
of vanishing gaugino masses. On the other hand, now both quark spins point in
the same direction, so that $J=1$. The produced squarks therefore have to be
in a $P-$wave, resulting in a cross section that is suppressed by two
additional powers of $\beta$ relative to $S-$wave reactions. Notice that for
the purpose of this distinction an $R-$type anti--squark acts like an $L-$type
squark, since it couples to a left--handed particle [$\overline{q_R} =
\left(\bar q\right)_L$]; similarly, an $L-$type anti--squark acts like an
$R-$type squark. Finally, pure $s-$channel processes also require the squarks
to be in a $P-$wave, i.e. have cross sections $\propto \beta^3$.

\begin{table}[t!] 
\begin{center}
\begin{tabular}{|r|c||c|c||c|c|c|c||c|} 
\hline
& & \multicolumn{2}{|c||}{diagrams} & helicity & thre-- &
\multicolumn{2}{|c||}{cross section [pb]} &  \\
No. & Process & QCD & EW & flip? & shold & QCD & QCD $+$ EW & ratio \\
\hline \hline
1 & $u u \rightarrow \ul \ul$ & $t, \, u$ & $t, \, u$ & yes & $\beta$& 0.683 &
0.794 & 1.162 \\
2 & $u u \rightarrow \ur \ur$ & $t, \, u$ & $t, \, u$ & yes & $\beta$& 0.761 &
0.796 & 1.045 \\
3 & $u u \rightarrow \ul \ur$ & $t, \, u$ & $t, \, u$ & no & $\beta^3$& 0.929 &
0.931 & 1.002 \\
4 & $d d \rightarrow \dl \dl$ & $t, \, u$ & $t, \, u$ & yes & $\beta$ & 0.198 &
0.232 & 1.171 \\
5 & $d d \rightarrow \dr \dr$ & $t, \, u$ & $t, \, u$ & yes & $\beta$ & 0.234 &
0.237 & 1.012 \\
6 & $d d \rightarrow \dl \dr$ & $t, \, u$ & $t, \, u$ & no & $\beta^3$ & 0.243
& 0.243 & 1.000 \\
7 & $u d \rightarrow \ul \dl$ & $t$ & $t, \, u$ & yes & $\beta$ & 0.969 & 1.22
& 1.261 \\ \hline &&&&&&&& \\[-1em] 
8 & $u \bar u \rightarrow \ul \ulb$ & $s, \, t$ & $s, \, t$ & no & $\beta^3$&
0.165 & 0.140 & 0.848
\\ 
9 & $u \bar u \rightarrow \ur \urb$ & $s, \, t$ & $s, \, t$ & no & $\beta^3$ &
0.187 & 0.170 & 0.909
\\ 
10 & $d \bar d \rightarrow \dl \dlb$ & $s, \, t$ & $s, \, t$ & no & $\beta^3$ &
0.0925 & 0.0784 & 0.847
\\ 
11 & $d \bar d \rightarrow \dr \drb$ & $s, \, t$ & $s, \, t$ & no & $\beta^3$ &
0.109 & 0.106 & 0.972
\\ 
12 & $u \bar u \rightarrow \dl \dlb$ & $s$ & $s, \, t$ & no & $\beta^3$ &
0.0341 & 0.0353 & 1.035 \\
13 & $d \bar d \rightarrow \ul \ulb$ & $s$ & $s, \, t$ & no & $\beta^3$ &
0.0207 & 0.0219 & 1.057 \\
14 & $u \bar d \rightarrow \ul \dlb$ & $t$ & $s, \, t$ & no & $\beta^3$ &
0.178 & 0.162 & 0.910 \\ \hline &&&&&&&& \\[-1em]
15 & $u d \rightarrow \ul \dr$ & $t$ & $t$ & no & $\beta^3$& 0.484 & 0.485 &
1.001 \\ 
16 & $u d \rightarrow \ur \dl$ & $t$ & $t$ & no & $\beta^3$ & 0.477 & 0.479 &
1.002 \\
17 & $u d \rightarrow \ur \dr$ & $t$ & $t$ & yes & $\beta$ & 1.113 & 1.114 &
1.001 \\
18 & $u \bar u \rightarrow \ul \urb$ & $t$ & $t$ & yes & $\beta$ & 0.569 &
0.569 & 1.000 \\ 
19 & $d \bar d \rightarrow \dl \drb$ & $t$ & $t$ & yes & $\beta$ & 0.331 &
0.331 & 1.000 \\
20 & $u \bar d \rightarrow \ul \drb$ & $t$ & $t$ & yes & $\beta$ & 0.491 &
0.491 & 1.000 \\
21 & $u \bar d \rightarrow \ur \dlb$ & $t$ & $t$ & yes & $\beta$ & 0.480 &
0.480 & 1.000 \\
22 & $u \bar d \rightarrow \ur \drb$ & $t$ & $t$ & no & $\beta^3$ & 0.202 &
0.203 & 1.004 \\
23 & $u \bar u \rightarrow \dr \drb$ & $s$ & $s$ & -- & $\beta^3$ & 0.0420 &
0.0421 & 1.002 \\
24 & $d \bar d \rightarrow \ur \urb$ & $s$ & $s$ & -- & $\beta^3$ &0.0240  &
0.0240 & 1.000 \\
\hline
\end{tabular}
\caption{The 24 different squark pair production processes involving first
  generation (s)quarks only; charge conjugate reactions are included
  in the cross section if they differ from the listed ones. The
  letters $s, \, t, \, u$ stand for the existence of $s-, \, t-$ and
  $u-$channel diagrams, respectively; this is listed separately for
  strong and electroweak interactions. We also list whether the
  exchange of a fermion in the $t-$ and/or $u-$channel requires a
  helicity flip. The fifth column describes the threshold behavior of
  the cross section, in terms of the squark velocity $\beta$ in the
  center--of--mass frame; a behavior $\propto \beta \ (\beta^3)$
  indicates an $S- \ (P-)$wave cross section. The values of the cross
  sections are for scenario SPS 1a (see Table~1). The last column
  shows the relative size of the electroweak contributions.}
\end{center}
\end{table}

With this understanding, let us come back to Table~2 and its three categories
of reactions. In the first category the electroweak contributions almost
always increase the cross section. The exception is $dd \rightarrow \dl \dr$,
where the interference term is negative; however, due to the small
hypercharges involved, the electroweak contribution is very small in this
case. In fact, as argued above, electroweak contributions are sizable only if
both produced squarks are $SU(2)$ doublets, although owing to the large
hypercharge the electroweak contribution to the $\ur \ur$ final state is not
completely negligible. The biggest electroweak contribution occurs in $ud
\rightarrow \ul \dl$. Here the gluino $t-$channel contribution interferes with
a chargino $u-$channel contribution; in the limit where one of the charginos
is a pure $SU(2)$ gaugino (charged wino), its couplings to a quark and a
squark exceed those of the corresponding neutralino by a factor $\sqrt{2}$.
One might therefore expect the electroweak contributions to this channel to be
twice as important as for $\ul \ul$ or $\dl \dl$ production. Table~2 shows
that the actual enhancement factor is somewhat smaller than this. The reason
is that the relative importance of the electroweak contributions to $\ul \ul$
and $\dl \dl$ production is enhanced by the destructive interference between
the two QCD gluino ($t-$ and $u-$channel) diagrams.

Note that all reactions in this category that can proceed via $SU(2)$
interactions require a helicity flip. In models with gaugino mass unification
the EW contributions are therefore significantly smaller than a naive guess
based on coupling constants only; this is only partly compensated by the fact
that the color factor for the interference terms is two times bigger than for
the pure QCD amplitudes, see e.g. Eq.(\ref{Cuu}).

The different absolute sizes of the total cross sections for these seven
reactions are determined by the interplay of four effects. Cross sections for
the production of two identical particles in the final state are suppressed by
the statistics factor $1/2$. On the other hand, we saw above that $\ul \ur$
and $\dl \dr$ production can only occur in a $P$ (or higher) partial wave; the
resulting cross sections are therefore suppressed by an extra factor of
$\beta^2$. In the given scenario this gives a similar suppression as the
statistics factor for producing two equal squarks; if squarks are heavier,
this threshold suppression over--compensates the statistics factor. Note also
that the flux of valence $u-$quarks in the proton is very roughly two times
larger than that of valence $d-$quarks. The ratio of the two pdf's increases
with increasing Bjorken$-x$; the preponderance of $\tilde u$ over $\tilde d$
squarks therefore becomes more prominent for larger squark masses. Finally, in
mSUGRA $SU(2)$ doublet squarks are somewhat heavier than their $SU(2)$ singlet
counterparts, since the masses of the latter are not enhanced by $SU(2)$
gaugino loop contributions. The pure QCD cross section for $SU(2)-$singlets is
therefore somewhat bigger than that for doublets. However, this difference is
largely canceled by the electroweak contributions, which, as we just saw, are
much more important for the doublets.

All seven reactions in the second category of Table~2 require an anti--quark in
the initial state; moreover, all cross sections suffer from $P-$wave
suppression. They are therefore substantially smaller than the cross sections
in the first group; this difference becomes even bigger for larger squark
masses. On the other hand, since the $\bar u$ and $\bar d$ densities in the
proton are similar, the cross sections for processes with $u$ quarks in the
initial state are now only about two times larger than those with $d-$quarks.

More important for our purposes is that for most of these processes,
electroweak contributions {\em reduce} the total cross section. This is
because both the interference between QCD $s-$channel and EW $t-$channel
diagrams, and that between QCD $t-$channel and EW $s-$channel diagrams is {\em
  destructive} for all these seven processes.\footnote{The relevant products
  of couplings, as well as the color factors, are positive; however, the
  remainder of the function $\Omega$ defined in Eq.(\ref{master}) is negative.
  On the other hand, QCD $s-$ and $t-$channel diagrams, where present,
  interfere constructively, due to an additional sign provided by the color
  factor, see e.e. Eq.(\ref{Cuubar}).}  This effect is most prominent for $q
\bar q \rightarrow \tilde q_L \bar{\tilde q}_L \ (q = u, d)$, where the
dominant EW contribution is due to $SU(2)$ gauge interactions. However, the EW
contribution is also sizable for $u \bar u \rightarrow \ur \urb$, due to the
relatively large hypercharges of the involved (s)quarks; recall that the
$SU(2)$ neutralino coupling gets a factor $1/2$ from weak isospin, which is
smaller than the hypercharge of the right--handed $u-$quark and its
superpartner. These processes do not require a helicity flip, i.e. there is no
factor of the mass of the exchanged fermion in the $t-$channel amplitudes. The
$t-$channel propagators then favor the lightest exchanged fermion. This
enhances the EW contributions relative to the QCD ones, and the $U(1)$
contribution relative to the one due to $SU(2)$ interactions. The relatively
mild suppression of the $\ul \dlb$ final state, which is also accessible via
$SU(2)$ interactions, can be explained from the observation that here no QCD
$s-$channel diagram contributes. This reduces the number of interference
terms, but does not change the pure QCD contribution much, since here
$s-$channel diagrams are subdominant.

Finally, $u \bar u \rightarrow \dl \dlb$ and $d \bar d \rightarrow \ul \ulb$
can proceed in QCD only through $s-$channel diagrams, but receive EW
contributions from $t-$channel (chargino exchange) diagrams. The interference
between these diagrams is again destructive. However, for these reactions this
is over--compensated by the squared EW $t-$channel contribution. The absence
of QCD $t-$channel diagrams makes the pure QCD and interference contributions
quite small. On the other hand, the factor $\sqrt{2}$ in each chargino
coupling relative to the $SU(2)$ neutralino coupling enhances the pure EW
$t-$channel contribution. For the case at hand, the pure QCD, pure EW and
interference contributions are of roughly equal absolute size; nevertheless,
because of the strong cancellation between the pure electroweak and
interference contributions, the total effect of the EW contributions only
amounts to a few percent. The total cross sections for these processes
therefore remain very small.

In the ten cases of the third category, which does not allow interference
between EW and QCD diagrams, the electroweak contributions are obviously
always positive, but very small. This is largely due to the fact that all
these reactions involve at least one $SU(2)$ singlet in the final state, so
that only $U(1)_Y$ interactions contribute. On the other hand, the cross
sections for the first eight ($t-$channel) processes are sizable; in fact,
$\ur \dr$ production is one of the most important channels also for heavy
squarks, since it can proceed from two valence quarks in the initial state,
and allows an $S-$wave in the final state. 

This discussion allows us to understand the small total size of EW
contributions due to hypercharge interactions. The processes of the first
category, which have relatively large cross sections, receive small, positive
corrections. In the second category we find (relatively) bigger, negative
corrections, but the cross sections are smaller; the total contribution from
$U(1)_Y$ interactions therefore remains positive. The relative importance of
these contributions is further reduced by the large number of processes that
receive only very small EW contributions, due to the absence of interference
with QCD diagrams and/or due to the small hypercharges of the involved
(s)quarks.

Coming back to Table~1, we saw that the relative importance of the electroweak
contributions increases with increasing gaugino to squark mass ratio.  This
can be explained from our earlier observation that the most important EW
contributions involve the interference of $t-$ and $u-$channel amplitudes
(category 1 in Table~2). As explained above, the amplitudes for all processes
of this kind that receive contributions from $SU(2)$ interactions are
proportional to a gaugino mass. These contributions are therefore sensitive to
the ratio of gaugino and squark masses. In mSUGRA the relative importance of
the EW contributions becomes largely insensitive to $m_{1/2}$ (for fixed
squark mass) once $m_{1/2} \gsim m_0$. The physical squark masses are then
essentially independent of $m_0$, i.e. $m_{\tilde q} \propto m_{1/2}$, so that
the ratios of gaugino and squark masses become independent of $m_{1/2}$. The
situation might be different in models without gaugino mass unification; we
will come back to this point shortly.

Finally, Table~1 also shows that the electroweak contributions become
relatively more important with increasing squark mass scale, although for
scenario SPS 2 this effect is over--compensated by the small ratio $m_{1/2} /
m_0$. Table~2 shows that nine out of the ten processes where no interference
occurs are suppressed by additional powers of the squark velocity $\beta$ in
the partonic cms frame, and/or by the presence of an anti--quark in the
initial state. Increasing the squark mass reduces the amount of available
phase space, i.e. the average $\beta$ is reduced.  Moreover, larger values of
Bjorken$-x$ are required, which reduces the flux of anti--quarks much more
strongly than that of quarks. For the group of five processes with
interference between $s-$ and $t-$channel diagrams where electroweak
contributions decrease the total cross section both suppression factors are
present. The relative importance of these processes therefore decreases
quickly with increasing squark mass. In contrast, the category 1 processes
which receive large, positive EW contributions all receive contributions from
two valence quarks in the initial state, and can have the squarks in an
$S-$wave. Increasing the squark masses therefore enhances the relative
importance of these processes.

\begin{figure}[h!] 
\begin{center}
\rotatebox{270}{\includegraphics[width=14cm]{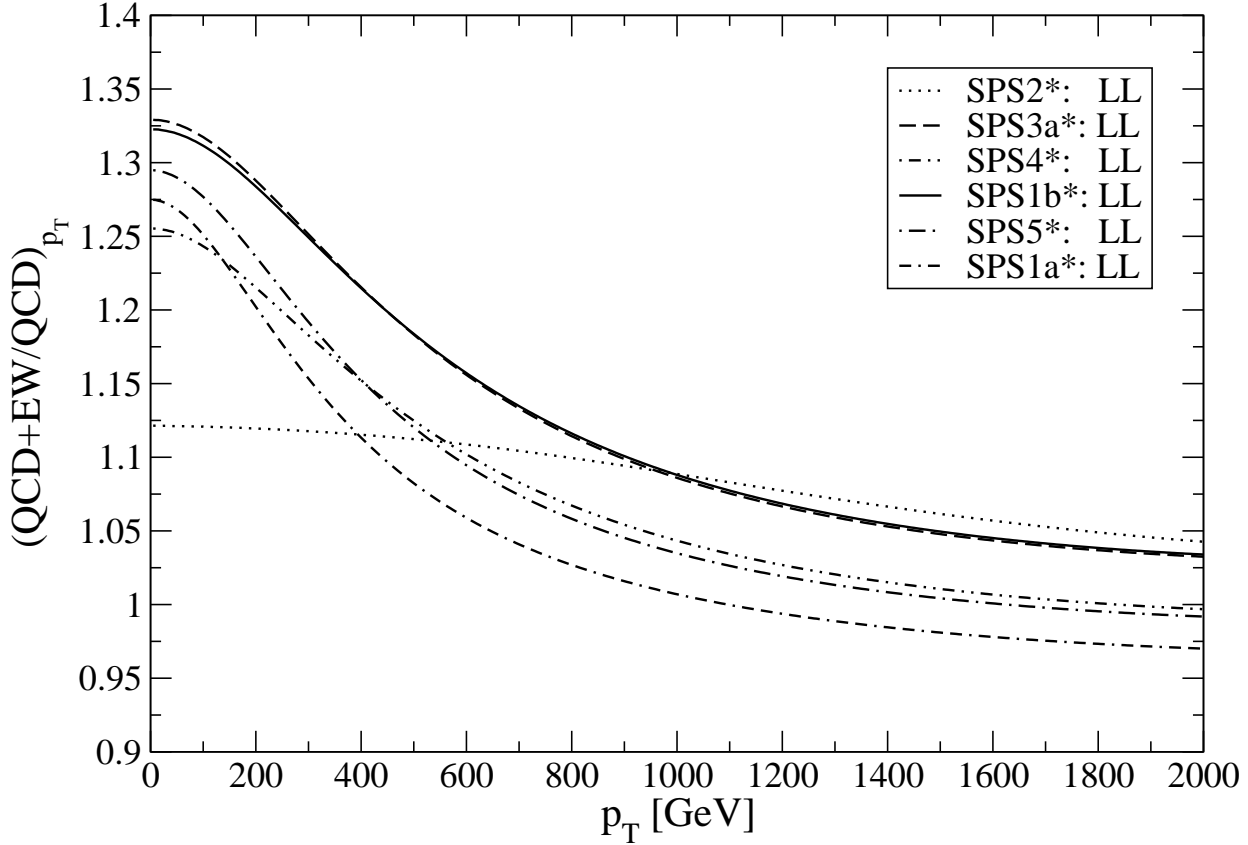}}
\caption{The ratio of QCD$+$EW to pure QCD predictions for the production of 
  two $SU(2)$ doublet (anti--)squarks at the LHC as a function of the squark
  transverse momentum. We use the same mSUGRA scenarios as in Table~1.}
\label{fig1}
\end{center}
\end{figure}

We emphasized above that the dominant EW contributions come from the
interference of $t-$ and $u-$channel diagrams with QCD diagrams. Since in
mSUGRA the electroweak gauginos are about three and six times lighter than the
gluino, one expects the EW contributions to be most prominent for small
transverse momenta of the produced squarks. This is borne out by
Fig.~\ref{fig1}, which shows the ratio of the tree--level differential cross
section with and without EW contributions. Here, and in the subsequent
figures, we concentrate on the production of two $SU(2)$ doublet
(anti--)squarks, where the EW contributions are largest.

The observed behavior can be understood from the interplay of several effects.
For simplicity assuming equal squark masses in the final state, the relation
between the partonic cms energy and squark transverse momentum can be written
as
\beq \label{ept}
\hat{s} = 4 \left( m^2_{\tilde q} + \frac{p_T^2 } {\sin^2 \theta} \right)\, ,
\eeq
where $\theta$ is the cms scattering angle. The parton flux in the initial
state is largest for smallest $\hat{s}$. Eq.(\ref{ept}) then shows that
configurations where $\sin^2\theta$ is maximal, i.e. where $\cos\theta$ is
small, are preferred if $p_T$ is sizable.

On the other hand, the denominators of the $t-$channel propagators can be
written as
\beq \label{eprop1}
\hat{t} - M^2_{\tilde V} = m^2_{\tilde q} -\frac{\hat{s}} {2} ( 1 - \beta
\cos\theta)  - M^2_{\tilde V} \, , 
\eeq
where $M_{\tilde V}$ is the mass of the exchanged gaugino; the expression for
$u-$channel propagators can be obtained by $\cos\theta \rightarrow
-\cos\theta$. These propagators therefore prefer large $\beta |\cos\theta|$;
however, $t-$ and $u-$channel propagators prefer different signs of
$\cos\theta$. We saw earlier that the dominant EW contributions are due to the
interference between $t-$ and $u-$channel diagrams (category 1 in Table~2).
These cross sections are proportional to a single power of the threshold
factor $\beta$. The steeply falling pdf's imply that these processes therefore
prefer rather small values of $\beta$ even for small $p_T$. As a first
approximation we can therefore ignore terms $\propto \beta \cos\theta$ in the
propagators. The ratio of EW and QCD $t-$ or $u-$channel propagators then
becomes
\beq\label{proprat}
\frac {\rm EW} {\rm QCD} = \frac {\hat{s}/2 - m^2_{\tilde q} + M^2_{\tilde g}
} {\hat{s}/2 - m^2_{\tilde q} + M^2_{\widetilde W}} \simeq
\frac {2 p_T^2 + m^2_{\tilde q} + M^2_{\tilde g}}
{2 p_T^2 + m^2_{\tilde q} + M^2_{\widetilde W}}\,,
\eeq
where $M_{\widetilde W}$ is the mass of the relevant chargino or neutralino.
Most of the mSUGRA scenarios of Table~1 and Fig.~\ref{fig1} have $m^2_{\tilde
  q} \sim M^2_{\tilde g} \gg M^2_{\widetilde W}$. Eq.(\ref{proprat}) shows
that the interference term will then be enhanced by a factor $\sim 2$ at small
$p_T$.  However, this enhancement disappears for $m^2_{\tilde q} \gg
M^2_{\tilde g}$, as in SPS 2.

Eq.(\ref{proprat}) shows that the propagator enhancement of the EW
contributions also disappears once $2 p_T^2 \gg m^2_{\tilde q}$. However, at
large $p_T$ the processes of category 1 may no longer be dominant even if we
insist on having two $SU(2)$ doublet (anti--)squarks in the final state. We saw
earlier that these processes require a helicity flip, i.e. their amplitudes
are proportional to the mass of the exchanged gaugino. Dimensional arguments
imply that the resulting product of two gaugino masses in these cross sections
has to be compensated by an extra factor of $p_T^{-2}$ for large $p_T$,
relative to the processes without helicity flip. Table~2 shows that only
category 2 processes can produce two doublet squarks without helicity flip.
Here the interference between QCD and EW diagrams is destructive. If these
processes dominate at large $p_T$, as implied by $p_T$ power counting, EW
contributions should therefore lead to a suppression of the cross section in
this region of phase space. However, we also saw above that all category 2
processes require an anti--quark in the initial state. The relevant parton
fluxes therefore fall much faster with increasing $\hat{s}$ than for some of
the category 1 processes. At large $p_T$ one therefore has two competing
suppression factors: category 1 processes, where EW contributions enhance the
cross section, are suppressed by an extra $p_T^{-2}$, while category 2
processes are suppressed by more quickly falling pdf's. Eq.(\ref{ept}) shows
that the latter suppression will be more relevant for larger squark masses.
Indeed, at large $p_T$ we observe the largest (or least negative) EW
contributions for scenarios with heaviest squarks. However, even in scenario
SPS1a, which has the smallest squark masses, EW contributions only suppress
the cross section by $\sim 3\%$ at large $p_T$; category 2 processes do
contribute significantly here, but do not really dominate.

\begin{figure}[h!] 
\begin{center}
\rotatebox{270}{\includegraphics[width=14cm]{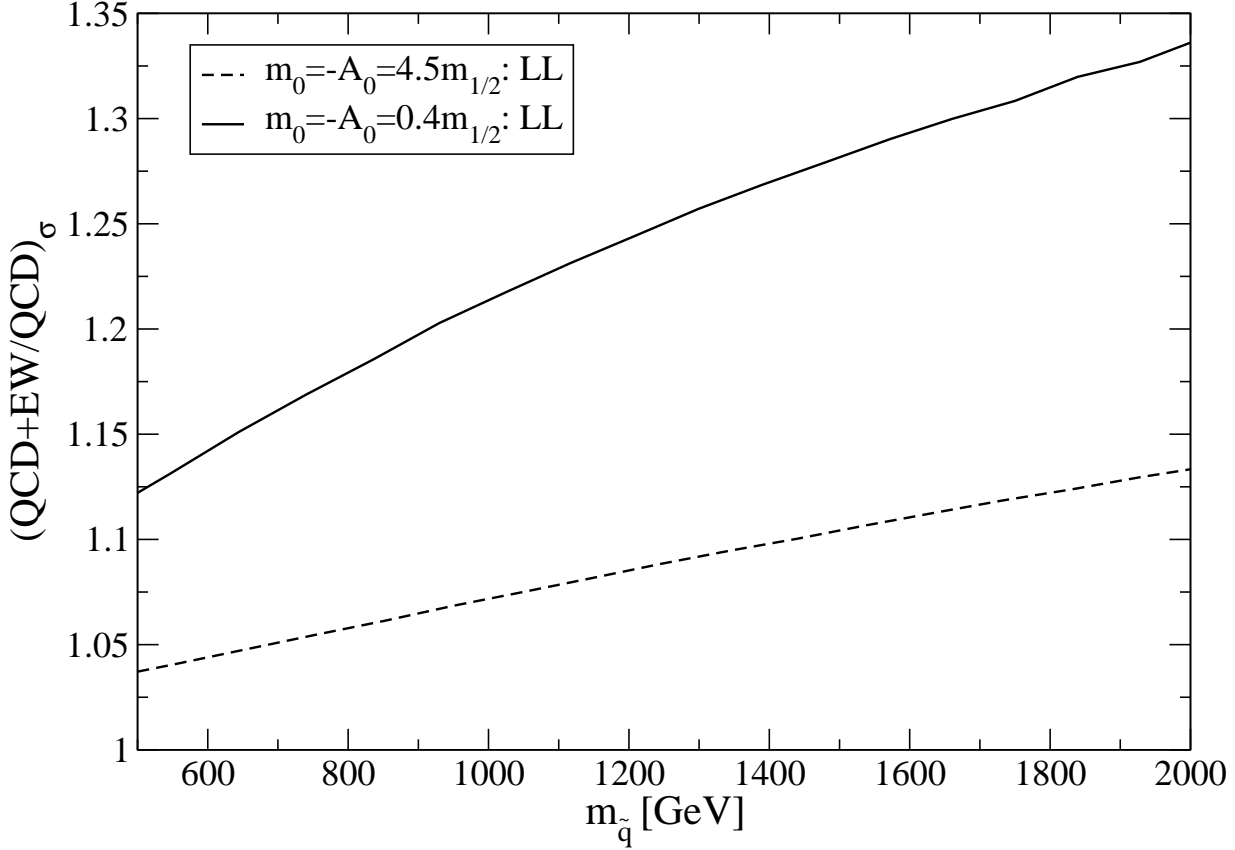} }
\caption{The ratio of QCD$+$EW to pure QCD predictions for the production of 
  two $SU(2)$ doublet (anti--)squarks at the LHC as a function of the squark
  mass. The upper (lower) curve is for $m_0 = 0.4 m_{1/2} \ (m_0 = 4.5
  m_{1/2})$, with the overall scale of these soft breaking parameters being
  varied.}
\label{fig2}
\end{center}
\end{figure}

We saw in Table~1 that the EW contributions tend to become more important with
increasing squark mass scale. This is further explored in Fig.~\ref{fig2},
which shows the ratio of the total cross section for the production of $SU(2)$
doublet squarks with and without EW contributions as function of the average
doublet squark mass. These curves have been generated by keeping the ratios of
the dimensionful mSUGRA input parameters $m_0, \, m_{1/2}$ and $A_0$ fixed,
but varying the overall mass scale; this corresponds to the ``benchmark
slopes'' of ref.\cite{sps}. We see that in a scenario with relatively large
gaugino masses, as in SPS 1a (upper curve), the EW contribution can increase
the cross section by more than 30\% for $m_{\tilde q} = 2$ TeV. The cross
section for much heavier squarks is too small for them to be detectable at the
LHC. A scenario\footnote{This curve includes a scenario very similar to SPS 2;
  however, the corresponding slope defined in \cite{sps} does not permit small
  squark masses.} with $m_0 = -A_0 = 4.5 m_{1/2}$ (lower curve) shows the same
trend; however, as noted earlier, the total EW contribution is much smaller in
this case, only reaching 13\% for $m_{\tilde q} = 2$ TeV.

Note that the maximal relative size of the EW contribution in Fig.~\ref{fig2}
exceeds that of the most favorable single process in Table~2. The results of
this Figure can therefore not be entirely due to the change of the relative
weights of the various processes, as described in our discussion of Table~1.
On top of that, the importance of the EW contributions to single processes
increases with increasing squark masses. This can be understood from the
behavior of the $t-$ and $u-$channel propagators. Smaller squark masses allow
larger values of $\beta$. The regions of phase space with large $|\cos
\theta|$ will then favor the squared $t-$ or $u-$channel propagators of pure
QCD contributions over the product of one $t-$ and one $u-$channel propagator
of the interference terms. This implies that increasing $m_{\tilde q}$ will
increase the relative importance of the interference terms relative to the
squared $t-$ and $u-$channel diagrams. This reduces the pure QCD contribution,
where the interference is destructive due to the negative color factor, see
e.g. Eq.(\ref{Cuu}), and enhances the importance of the EW contributions.

Comparison of the two curves in Fig.~\ref{fig2} reinforces the importance of
the gaugino masses. So far we have considered sparticle spectra generated with
mSUGRA boundary conditions. In particular, this implies that $SU(2)$ and
$U(1)_Y$ gauginos are much lighter than gluinos. Since the dominant EW
contributions (from category 1 in Table~2) are proportional to the product of
the gluino mass with the mass of an electroweak gaugino, we expect that these
contributions are sensitive to the assumed ratio of gaugino masses.

\begin{figure}[h!] 
\begin{center}
\rotatebox{270}{\includegraphics[width=14cm]{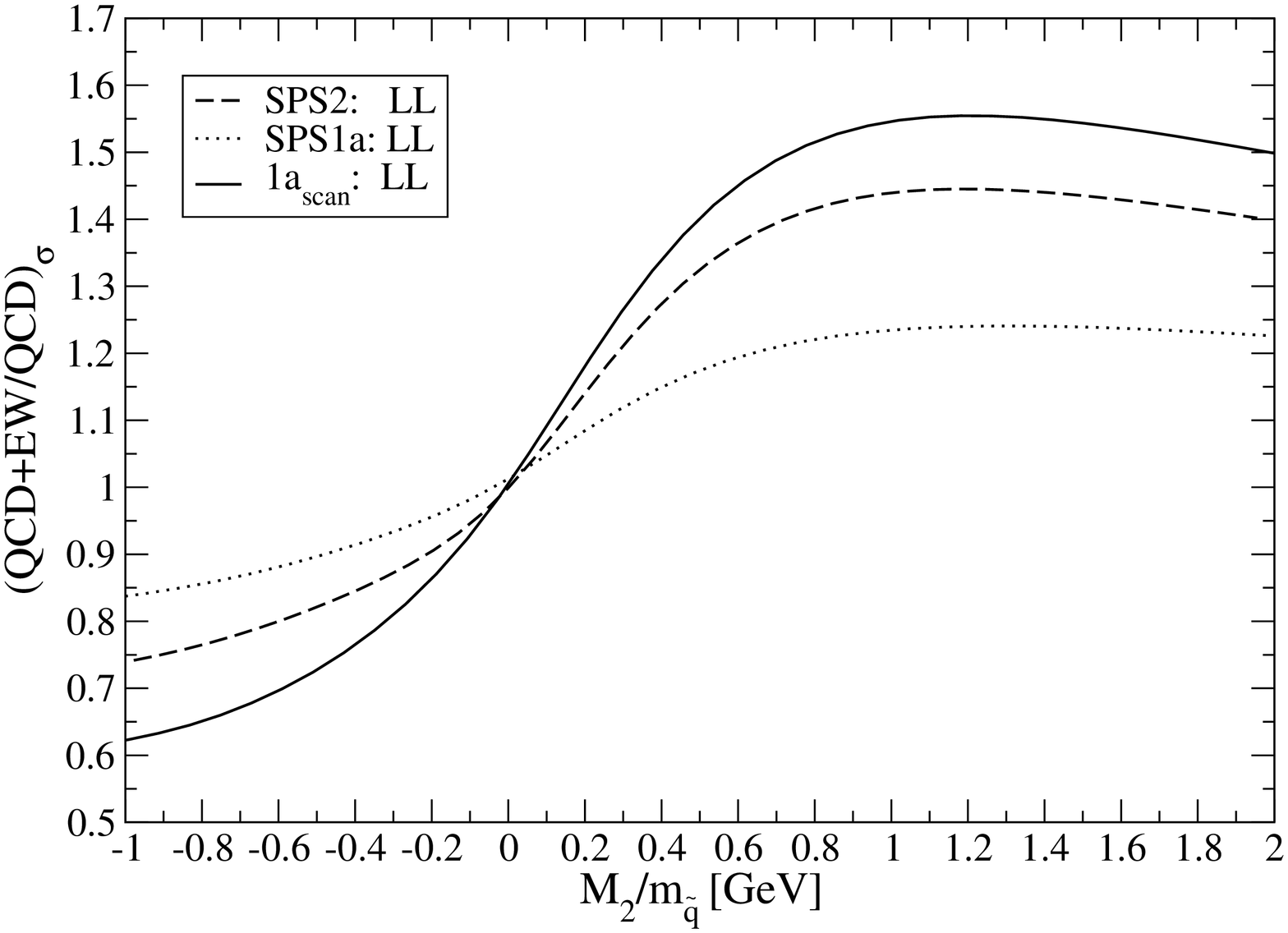} }
\caption{The ratio of QCD$+$EW to pure QCD predictions for the production of 
  two $SU(2)$ doublet (anti--)squarks at the LHC as a function of the ratio of
  the $SU(2)$ gaugino mass parameter $M_2$ and the squark mass. The solid and
  dotted curves are both based on scenario SPS 1a of Table~1, but for the
  solid curve all soft breaking masses have been scaled up to achieve a squark
  mass of 2 TeV. The dashed curve is for scenario SPS 2. In all cases $M_2$
  has been varied directly at the weak scale using SPheno \cite{spheno},
  leaving all other weak--scale soft breaking parameters unchanged.}
\label{fig3}
\end{center}
\end{figure}

This is demonstrated in Fig.~\ref{fig3}, where we vary the $SU(2)$ gaugino
mass $M_2$ at the weak scale, keeping all other parameters fixed. We see that
the electroweak contributions become maximal if $M_2 \simeq m_{\tilde
  q}$.\footnote{The phenomenology of models with large $SU(2)$ gaugino mass
  parameter has recently been discussed in \cite{bmst}; however, EW
  contributions to squark pair production were not considered in that paper.}
This can be understood from the observation that this choice maximizes $M_2 /
|\hat{t} - M_2^2|$, see Eq.(\ref{eprop1}). In a scenario with $m_{\tilde g}
\simeq m_{\tilde q}$ and large squark mass (solid curve), this can lead to EW
contributions in excess of 50\%. In scenario SPS 2 (dashed curve) the
contributions remain somewhat smaller, partly because of the reduced squark
mass, and partly because the lower gluino mass reduces the importance of the
interference terms. Not surprisingly, taking $m_{\tilde g} \simeq m_{\tilde
  q}$ also maximizes the size of those pure QCD contributions that require a
helicity flip.\footnote{Maximizing the cross section with respect to $M_2$ and
  $m_{\tilde g}$ would prefer $m_{\tilde g}$ slightly below $M_2$. The reason
  is that, as emphasized repeatedly, the main EW effect is due to interference
  between $u-$ and $t-$channel diagrams. The corresponding mixed product of
  propagators has no strong preference for large $|\cos\theta|$. We saw above
  that, as a result, the denominator of the propagators are only moderately
  enhanced if the mass of the exchanged fermion is reduced. In contrast, the
  dominant QCD contributions come from squared $t-$ and $u-$channel diagrams.
  Here the propagators do show stronger preference for sizable $|\cos\theta|$,
  and hence to a smaller mass of the exchanged fermion.}  Finally, in scenario
SPS 1a with its relatively light squarks (dotted curve) the EW contribution
never goes much beyond 20\%. We saw in Table~2 that in this case the processes
of category 2 still contribute significantly. Here the EW contributions are
negative. Since these processes do not require a helicity flip, the absolute
size of the EW contributions decreases monotonically with increasing $|M_2|$.
As a result, the dotted curve reaches its maximum for somewhat larger values
of $M_2$; moreover, the maximum is less pronounced.

In Fig.~\ref{fig3} we show results as function of the weak scale soft breaking
parameter $M_2$, normalized to the squark mass. This parameter can be
negative. If we keep the sign of the gluino mass parameter positive, the sign
of the $t-u$ interference terms, which require a helicity flip, will
change.\footnote{One can write the amplitudes also entirely in terms of
  physical, positive masses. A negative $M_2$ would then need an extra factor
  $i$ for each coupling of an $SU(2)$ gaugino, giving an extra $i^2 = -1$ for
  the $SU(2)$ amplitude relative to the QCD amplitude.} We see that taking
$M_2$ large and negative will lead to cross sections that are significantly
reduced from the pure QCD contribution. The relative size of the EW
contributions is slightly smaller than that for positive $M_2$. This is partly
because we did not change the sign of the $U(1)_Y$ gaugino mass, keeping the
corresponding contribution positive (but very small). Moreover, the cross
sections for the (subdominant) category 2 processes remain essentially
unchanged when the sign of $M_2$ is flipped; recall that category 3 processes
do not contribute here. Altogether we see that the total cross section for the
production of two $SU(2)$ (anti--)squarks can change by up to a factor 2.5 as
$M_2$ is varied between $-m_{\tilde q}$ and $m_{\tilde q}$, if squarks are
quite heavy and $m_{\tilde g} \simeq m_{\tilde q}$.

As discussed earlier, EW contributions will be much smaller if at least one
(anti--)squark in the final state is an $SU(2)$ singlet. However, it might
well be possible to experimentally separate these different classes of final
states. At least for $m_{\tilde g} \gsim m_{\tilde q} > |M_2|, |M_1|$ the
production of two doublet squarks leads to significantly different final
states than that of singlet squarks. The former prefer to decay into charginos
and neutralinos dominated by $SU(2)$ gaugino components \cite{bbkt}; in
mSUGRA, this typically means $\tilde \chi_2^0$ and $\tilde \chi_1^\pm$. This
leads to longer decay chains than for $SU(2)$ singlet squarks, which prefer to
decay into the neutralino with the largest bino component \cite{bbkt}, in
mSUGRA usually $\tilde \chi_1^0$. The contribution of doublet squarks can
therefore be enhanced experimentally by requiring the presence of energetic,
isolated charged leptons (electrons or muons), in addition to $\geq 2$ jets
and missing transverse momentum \cite{nojiri}. If $m_{\tilde g} < m_{\tilde
  q}$ this distinction becomes more difficult, since then most squarks decay
into a gluino and a quark; however, the branching ratio for $SU(2)$ doublet
squark decays into charginos and neutralinos remains sizable even in this
case. Similarly, in models with explicit violation of $R-$parity, $SU(2)$
doublet squarks may have very different decay channels than singlets
\cite{herbi_new}.

\section{Summary and Conclusions}

In this paper we analyzed electroweak (EW) contributions to the production of
two squarks or anti--squarks at the LHC. We provided explicit expressions for
the squared matrix elements for all processes with (anti--)quarks in the
initial state, allowing for different squark masses in the final state. Not
surprisingly, corrections due to $SU(2)$ interactions are more important than
those from $U(1)_Y$ interactions. In both cases the dominant effect is from
the interference of electroweak and QCD interactions. The sign of the
interference between EW $t-$ and QCD $u-$channel diagrams (or vice versa) is
positive (negative) for equal (opposite) signs of the electroweak and QCD
gaugino mass parameters. Interference between EW $t-$ and QCD $s-$ channel
diagrams (or vice versa) is usually negative, and independent of the sign of
the gaugino mass parameters.

The physical significance of our results is threefold:
\begin{itemize}
  
\item The EW contributions can change the total cross section significantly.
  Focusing on the production of two $SU(2)$ doublet ($L-$type) squarks, we
  found the contributions with interference between $t-$ and $u-$channel
  diagrams to be dominant. For squark masses near the discovery reach of the
  LHC, EW effects can reduce or enhance the total cross section by more than a
  factor 1.5, if the absolute value of the $SU(2)$ gaugino soft breaking mass
  is near $m_{\tilde q}$; even in scenarios with gaugino mass unification the
  EW contribution can still change the cross section for the production of two
  $SU(2)$ doublet squarks by more than a factor 1.3. Recall that $SU(2)$
  doublet squarks often lead to different final states than singlet squarks
  do, allowing to distinguish these modes experimentally.
  
\item The EW contributions might give a new, independent handle on the gaugino
  mass parameters. In particular, we just saw that they are sensitive to
  relative {\em signs} between gaugino mass parameters, which might be
  difficult to determine using kinematical distributions only. For example, in
  anomaly--mediated supersymmetry breaking \cite{amsb} the products of
  electroweak and QCD gaugino masses are negative. In order to realize this
  potential, both the experimental and the theoretical uncertainties should be
  reduced to the 10\% level. This is certainly challenging, but should
  eventually be possible if squarks are not too heavy.
  
\item The EW contributions allow production of two squarks without color
  connection between the squarks. In contrast, at least in leading order QCD
  diagrams for squark production, there is color flow between the two squarks.
  The absence of such a flow could in principle give rise to a rapidity gap,
  i.e. a rapidity region into which no QCD radiation is emitted; such
  radiation would then only occur in the region between the squark and the
  quark from which it was produced (and from which it inherited its color).
  This is completely analogous to the gap predicted for ordinary
  (non--supersymmetric) two--jet events produced at hadron colliders via
  electroweak interactions \cite{zepgap}.  We are currently investigating if
  such supersymmetric rapidity gap events might be detectable at the LHC.

\end{itemize}

\subsection*{Acknowledgments}
This work was partially supported by the Bundesministerium f\"ur Bildung und
Forschung under contract no. 05HT6PDA. SB thanks the
Universit\"atsgesellschaft Bonn -- Freunde, F\"orderer, Alumni e.V. for
financial support.

\begin{appendix}
\section{Couplings}

In this Appendix we give explicit expressions for all the couplings appearing
in Sec.~2, using the notation of \cite{gh}. We only list couplings of
squarks. The corresponding couplings of anti--squarks can be obtained using 
relations (\ref{swapuu}) and (\ref{swapud}).

\subsection{Neutralino and Gluino Couplings}

Since gluino and neutralino exchange always occur together, we labeled them
with the subscript $l$, with $l \in \{1,2,3,4\}$ denoting the $l-$th
neutralino mass eigenstate and $l=5$ denoting the gluino. Here we need the
left-- and right--handed quark--squark--gaugino couplings, generically denoted
by $a$ and $b$, respectively. The relevant {\bf neutralino couplings} are:
\begin{eqnarray} \label{neutcoup}
a_{\tilde \chi^0_l}(\tilde d_{i\alpha}) &=& -\delta_{1\alpha} \sqrt{2} g 
\left( \frac{1}{2} N^{\ast}_{l2} - \frac{1}{6} \tan\theta_W N_{l1}^{\ast}
\right)\,,
\nonumber\\
b_{\tilde \chi^0_l}(\tilde d_{i\alpha}) &=& \delta_{2\alpha} 
\frac{\sqrt{2}}{3} g \tan\theta_W N_{l1}\,,
\nonumber\\  
a_{\tilde \chi^0_l}(\tilde u_{i\alpha}) &=& \delta_{1\alpha} \sqrt{2} g 
\left( \frac{1}{2} N_{l2}^{\ast} + \frac{1}{6} \tan\theta_W N_{l1}^{\ast}
\right)\,,
\nonumber\\
b_{\tilde \chi^0_l}(\tilde u_{i\alpha}) &=& -\delta_{2\alpha}
\frac{2\sqrt{2}}{3} g \tan\theta_W N_{l1}\,.
\end{eqnarray}
Here $\alpha = 1 \ (2)$ stands for an $L- \ (R-)$type squark, and $N_{l1}$ and
$N_{l2}$ stand for the $U(1)_Y$ (bino) and $SU(2)$ (neutral wino) components
of $\tilde \chi_l^0$, respectively. Recall that we ignore quark mass effects,
and hence also Yukawa contributions to the neutralino couplings. Finally, $g$
is the $SU(2)$ gauge coupling, and $\theta_W$ is the weak mixing angle.

The corresponding {\bf gluino couplings} are:
\begin{eqnarray} \label{gluinocoup}
a_{\tilde g}(\tilde d_{i\alpha}) &=& -\delta_{1\alpha} g_s \sqrt{2}\,,
\nonumber\\
b_{\tilde g}(\tilde d_{i\alpha}) &=& \delta_{2\alpha} g_s\sqrt{2}\,,
\nonumber\\
a_{\tilde g}(\tilde u_{i\alpha}) &=& -\delta_{1\alpha} g_s\sqrt{2}\,,
\nonumber\\
b_{\tilde g}(\tilde u_{i\alpha}) &=& \delta_{2\alpha} g_s \sqrt{2}\,,
\end{eqnarray}
where $g_s$ is the $SU(3)$ gauge coupling.

\subsection{Chargino Couplings}

For a given process, charginos cannot be exchanged in diagrams of the same
topology as neutralinos and gluinos. Here the subscript $l$ labeling the
exchanged particle therefore only runs from 1 to 2, corresponding to the two
chargino mass eigenstates in the MSSM. Their relevant couplings are:
\begin{eqnarray} \label{charcoup}
a_{\chi^+_l}(\tilde d_{i\alpha}) &=& -g U_{l1} \delta_{1\alpha}\,,
\nonumber\\
b_{\chi^+_l}(\tilde d_{i\alpha}) &=& 0\,,
\nonumber\\
a_{\chi^+_l}(\tilde u_{i\alpha}) &=& -g V_{l1} \delta_{1\alpha}\,,
\nonumber\\
b_{\chi^+_l}(\tilde u_{i\alpha}) &=& 0\,.
\end{eqnarray}
The vanishing of the right--handed, $b-$type couplings is again due to our
neglect of Yukawa couplings.

\subsection{Gauge Boson Couplings}

Here ($s-$channel) diagrams with photon, $Z$ boson and gluon exchange always
occur together. We therefore labeled these particles with subscript $l$,
$l=1,2,3$ standing for the $\gamma$, $Z$ boson and gluon, respectively.  $e$
and $f$ represent the left-- and right--handed gauge boson--quark--anti-quark
couplings,
\begin{eqnarray} \label{coupef}
e_{\gamma}(d_i,\bar d_i) &=& f_{\gamma}(d_i,\bar d_i) = \frac{1}{3} g
\sin\theta_W \,,
\nonumber\\
e_{\gamma}(u_i,\bar u_i) &=& f_{\gamma}(u_i,\bar u_i) = - \frac{2}{3} g
\sin\theta_W \,,
\nonumber\\
e_Z(d_i,\bar d_i) &=& \frac {g} {\cos\theta_W} \frac{1}{2} \left( 1 -
  \frac{2}{3} \sin^2\theta_W \right) \,,
\nonumber\\ 
f_Z(d_i,\bar d_i) &=& -\frac{g}{\cos\theta_W} \frac{1}{3} \sin^2\theta_W\,,
\nonumber\\
e_Z(u_i,\bar u_i) &=& -\frac{g}{\cos\theta_W} \frac{1}{2} \left(
1 - \frac {4}{3} \sin^2\theta_W \right)\,,
\nonumber\\
f_Z(u_i,\bar u_i) &=& \frac{g}{\cos\theta_W} \frac{2}{3} \sin^2\theta_W\,,
\nonumber\\
e_g(d_i,\bar d_i) &=& f_g(d_i,\bar d_i) = e_g(u_i,\bar u_i) = f_g(u_i,\bar
u_i) = -g_s\, .
\end{eqnarray}
Their couplings to a squark and an anti--squark, generically denoted by $c$,
are:
\begin{eqnarray} \label{coupc}
c_{\gamma}(\tilde d_{i\alpha},\bar{\tilde d}_{i\alpha}) &=& \frac{1}{3} g
\sin\theta_W \,,
\nonumber\\
c_{\gamma}(\tilde u_{i\alpha},\bar{\tilde u}_{i\alpha}) &=& -\frac{2}{3}
\sin\theta_W \,,
\nonumber\\
c_Z(\tilde d_{i\alpha},\bar{\tilde d}_{i\alpha}) &=& \frac{g}{\cos\theta_W}
\frac{1}{2} \left( \delta_{1\alpha} - \frac{2}{3} \sin^2\theta_W \right) \,,
\nonumber\\
c_Z(\tilde u_{i\alpha},\bar{\tilde u}_{i\alpha}) &=& \frac{g}{\cos\theta_W} 
\frac{1}{2} \left( -\delta_{1\alpha} + \frac{4}{3} \sin^2\theta_W \right)\,,
\nonumber\\
c_g(\tilde d_{i\alpha},\bar{\tilde d}_{i\alpha}) &=& c_g
(\tilde u_{i\alpha},\bar{\tilde u}_{i\alpha}) = -g_s \,.
\end{eqnarray}
Note that in the absence of $L-R$ mixing, the couplings listed in
Eq.(\ref{coupc}) are nonzero only if both the squark and the anti--squark are
$SU(2)$ doublets ($\alpha= 1$), or both are singlets ($\alpha=2$).

Finally, in some cases there are $s-$channel diagrams in which a $W-$boson is
exchanged. Its couplings to the initial state are given by
\begin{eqnarray} \label{coupW1}
e_W(d_i,\bar u_i) &=& e_W(u_i,\bar d_i) = -\frac{g}{\sqrt{2}}\,,
\nonumber\\
f_W(d_i,\bar u_i) &=& f_W(u_i,\bar d_i) = 0\,,
\end{eqnarray}
and the relevant final state couplings are
\begin{eqnarray} \label{coupW2}
c_W(\tilde d_{i\alpha}, \bar{\tilde u}_{i\alpha}) &=&
c_W(\tilde u_{i\alpha},\bar{\tilde d}_{i\alpha}) = - \frac{g}{\sqrt{2}}
 \delta_{1\alpha} \, . 
\end{eqnarray}

\end{appendix}

\end{document}